\documentclass{aa}

\usepackage{natbib,twoopt}
\usepackage[breaklinks=true]{hyperref}

\usepackage[varg]{txfonts}
\usepackage{graphicx}

\bibpunct{(}{)}{;}{a}{}{,}
\hypersetup{
    colorlinks,
    linkcolor=magenta,
    citecolor=blue,
    urlcolor=red}

\begin{document}

\title{Oscillations of red giant stars with magnetic damping in the core}
\subtitle{II. Mixed mode visibilities on the red-giant branch}

\author{Jonas M\"uller\inst{1,2}
\and Saskia Hekker\inst{1,2}}

\institute{Heidelberger Institut für Theoretische Studien, Schloss-Wolfsbrunnenweg 35, 69118 Heidelberg, Germany
\and Zentrum für Astronomie (ZAH/LSW), Heidelberg University, Königstuhl 12, 69117 Heidelberg, Germany}

\titlerunning{Oscillations of red giant stars with magnetic damping in the core: II.}
\authorrunning{M\"uller \& Hekker}

\date{Received <date> /Accepted <date>}

\abstract{The visibility of oscillation modes can be used to draw conclusions about their average energy. Visibilities can be estimated from observed power spectra or from theory by making assumptions about the damping processes occurring in the star. However, a quantitative comparison between the two approaches was so far not feasible due to observational biases. The biases arise from the fact that in observations, the power spectrum is divided into frequency segments in which modes of a certain spherical degree are expected to dominate. This can lead to modes that correspond to the spherical degree under consideration being neglected, and modes of a different degree contributing to the observed visibility.
In this work, we used synthetic power spectra to calculate the visibility as it has been done in observations and compare it with published observed visibilities to quantify the influence of the biases. We find that, taking the biases into account, the observed spatial response of the dipole modes is 1.47, which is closer to the theoretical value than previous estimates.
In particular, we predict that the normalized dipole mode visibility of late red-giant branch (RGB) stars might be overestimated by up to 20\% in published observations. For stars with depressed dipole modes, we find that the normalized dipole mode visibilities estimated in observational studies might be overestimated by ${\sim}$20\% throughout their entire evolution on the RGB.
The quadrupole mode visibility, on the other hand, appears to be largely unaffected by the biases, expect on the late RGB.
In addition, we investigated the evolution of the visibility and detectability of the mixed mode signature while testing different prescriptions for the energy loss caused by a strong internal magnetic field in the stellar core. We argue that taking into account the inner turning point of the g-mode cavity could allow a portion of the mode energy to be preserved when interacting with a strong magnetic field. We further show that such partial dissipation allows the mixed mode signature to be both present or absent in the observable power spectra, which is consistent with observations of stars with depressed dipole modes.}


\maketitle

\section{Introduction} \label{sect: Introduction}

Asteroseismology has benefited greatly from the wealth of high-quality stellar light curves provided by observations from space-based telescopes such as {\it Kepler} \citep{borucki+10}. The light curves can be converted into power spectral densities (PSDs), which have been intensively analyzed using a variety of different approaches \citep[e.g.,][]{appourchaux+08,stello+09,bedding+10,hekker+10,kallinger+10}. In the PSDs of red-giant branch (RGB) stars, mixed multipolar modes have been discovered \citep{deRidder+09}. These modes are of particular interest because they behave like a gravity mode ("g-mode") in the core and like a pressure mode ("p-mode") in the outer layers of the star. We refer to the regions of the star in which the oscillations resonate as “cavities.” In RGB stars, the p- and g-mode cavity are separated by an intermediate layer in which the oscillations decay, known as the “evanescent zone." Mixed modes can be used to directly probe the interior of RGB stars, which has led to a number of groundbreaking discoveries \citep[see e.g.,][and the references therein]{chaplin+miglio13,hekker+17,garcia+ballot19}.

An important property of the oscillation modes is their visibility. The mode visibility is defined as the ratio of the total squared modulus of the amplitude of the multipole modes of a given spherical degree $\ell > 0$ to that of the radial modes with $\ell = 0$. Since the mode amplitude is a proxy for the average mode energy, the visibility can be used to investigate the energetics of the modes. From a theoretical point of view, the total squared modulus of the amplitude of the modes of degree $\ell$ can be estimated by integrating the power spectrum corresponding to this $\ell$ over the frequency range under consideration, which typically consists of a few pressure radial orders. 
However, in observations it is not possible to completely separate the contributions of the different $\ell$ to the overall power spectrum. Therefore, most authors have divided the frequency range under consideration into segments in which modes of a certain $\ell$ are mainly expected to be observable, over which they then integrate the power spectrum \citep[e.g.,][]{mosser+12, stello+16a, stello+16b, mosser+17, dreau+21}.
This can lead to a bias in the observed visibilities, due to the spread of the gravity-dominated mixed modes away from the position of their nominal p-modes, placing them outside the frequency segments selected for this $\ell$. Thus, these modes are neglected in the calculation of the visibility \citep{mosser+12}. Following this line of reasoning, these gravity-dominated mixed modes could instead contribute to the visibility of another $\ell$, which is then increased beyond its actual value. Therefore, theoretical and observable visibilities should not be directly compared with each other.

Additional aspects that complicate the comparison of the theoretical and observable visibility are the $\ell$-dependent effects that depend on the observation technique and geometric effects, such as limb darkening \citep{ballot+11}, which we refer to as the spatial response, $\mathcal{S}_{\ell}$\footnote{Another aspect that influences the observable visibility is the bell-shaped power excess of the modes in the PSD of RGB stars. However, this contribution is not included in $\mathcal{S}_{\ell}$ and is considered separately.}.
If the total energy of the modes is distributed evenly across all $\ell$ (i.e., energy equipartition), the observable visibilities should scatter around the theoretical value of $\mathcal{S}_{\ell}$. However, several observational studies \citep[e.g.,][]{mosser+12,stello+16a,stello+16b} have found that the observed visibilities of the dipole modes tend to be lower than expected, while the visibilities of the quadrupole modes appear to be higher. If this is true, energy equipartition may not be realized.
On the other hand, \citet{mosser+12} suggested that these offsets could potentially be explained, at least in part, by the bias mentioned in the previous paragraph.
In this scenario, the absence of dipole mode energy could be due to the spread of the gravity-dominated mixed dipole modes away from the location of the nominal p-mode, causing them to be neglected in the calculation of the visibility of the dipole modes. These gravity-dominated dipole modes could then instead contribute to the visibility of the quadrupole modes, thus increasing their observed visibility compared to the theoretical value.
However, this has not yet been rigorously tested.

We refer to RGB stars with a normal multipole mode visibility as “normal stars.” Within the sample of RGB stars observed by {\it Kepler}, there is a subgroup with unusually low multipole mode visibilities \citep{mosser+12,stello+16a,stello+16b}. Due to the low amplitudes of their dipole modes, we refer to these stars as  the "depressed dipole mode stars"\footnote{In contrast to the first article in this series, we use the term “depressed” instead of “suppressed” to distinguish cases in which the energy is the g-mode cavity is partly dissipated (i.e., "depressed") and completely dissipated (i.e., "suppressed") \citep[see][]{mosser+17}.}. The low multipole mode amplitudes of the depressed dipole mode stars imply that an efficient damping mechanism is at work in their cores \citep[e.g.,][]{garcia+14, fuller+15, coppee+24}.
\citet{fuller+15} suggest that the additional damping might be caused by the interaction of internal gravity waves with a strong magnetic field in the stellar core. Incident gravity waves could scatter into slow magnetic waves that propagate upward and are likely to dissipate quickly, causing a complete loss of mode energy in the g-mode cavity \citep{lecoanet+17,rui+fuller23,david+26}\footnote{However, \citet{rui+fuller23} point out that a global solution is required to fully characterize the structure of the modes.}. Crucially, this implies that the multipole modes of the depressed dipole mode stars cannot be mixed and their amplitudes are determined by the amount of energy that does not leak into the core.
While the absence of the mixed mode signature seems to correspond to some depressed dipole mode stars, \citet{mosser+17} show that the dipole modes of a significant fraction of these stars are mixed. This means that, at least for a subset of the depressed dipole mode stars, the damping mechanism must allow a small amount of energy to escape from the g-mode cavity and to contribute to the observable oscillations.
In this context, ray tracing studies of magneto-gravity waves predict that the interaction of the waves with an internal magnetic field induces only partial dissipation of the mode energy \citep{loi20a,loi20b,muller+25}, which allows for the presence of mixed modes in the PSDs of the depressed dipole mode stars.

Magnetic fields can occur not only in the core of red giants but also in the envelope. Observational studies of stellar activity in red giants suggest that surface magnetic fields can inhibit the excitation of their oscillation modes \citep{gaulme+20,gehan+22,gehan+24,crawford+25}. Although surface magnetic fields can thus also reduce the amplitudes of the mixed modes, it is not related to the occurrence of the depressed dipole mode stars. This is because surface magnetic fields also reduce the amplitudes of the radial modes, leaving the visibility of the mixed modes unaffected. Therefore, we do not consider surface magnetic fields in this study.

Here, we generate synthetic PSDs for numerical models of RGB stars of different masses using the procedure of \citet{muller+26}, which allows for the treatment of mode energy loss in the p- and g-mode cavity. In Sect. \ref{SECT: Methods}, we describe the generation of synthetic PSDs and the computation of both the theoretical and observable visibility from these spectra. With this setup, we pursue three main objectives.
First, in Sect. \ref{SECT: spatial response}, we use publicly available data from \citet{stello+16a,stello+16b} to calculate an estimate of the spatial response, $\mathcal{S}_{\ell}$, for the dipole and quadrupole modes, taking into account the biases caused by integrating the power spectrum over $\ell$-dependent segments of the frequency range. Then, we compare our estimates to the theoretical values of $\mathcal{S}_{\ell}$ to check whether energy equipartition is realized.
Second, in Sect. \ref{SECT: Theoretical versus observable visibilities}, we investigate how well theoretical and observable visibilities can be compared throughout the evolution of the stars along the RGB, assuming different efficiencies of the damping process in the g-mode cavity.
Third, in Sect. \ref{SECT: Observable signatures of magnetic damping}, we test three implementations of the energy loss caused by a strong internal magnetic field and compute the resulting dipole and quadrupole visibilities as the stars ascend the RGB. In addition, we investigate how the energy loss affects the detectability of the mixed mode signature.
Lastly, we conclude in Sect. \ref{SECT: Conclusions}.

\section{Methods}  \label{SECT: Methods}

\subsection{Synthetic power spectrum} \label{sect: power spectrum}

\citet{muller+26} developed an analytical function that approximately expresses the power spectrum of a solar-like oscillator as a function of the oscillation frequency, $\nu$, up to a proportionality factor. This function can be used to generate synthetic PSDs that take into account the damping processes in both the p- and g-mode cavity. To do this, we first introduce the normalized power spectrum $P_\ell$\footnote{In the article by \citet{muller+26}, the normalized power spectrum is denoted by $p$. Here, $p$ stands for the pressure.} of the spherical degree $\ell$ and the quantities it depends on:
\begin{gather}
   P_\ell = \left(1 + e^{-4\Im(\Phi_{\rm g})}\right) \frac{A}{\mathcal{H}_{\rm p}}.
   \label{eq: normalized power spectrum}
\end{gather}
Here, $\Im(\Phi_{\rm g})$ denotes the imaginary part of $\Phi_{\rm g}$. The parameter $\Phi_{\rm g}$ is a frequency-dependent complex phase that governs the influence of the g-mode cavity on the observable oscillations. It can be expressed as 
\begin{gather}
    \Phi_{\rm g} = \arctan\left(q_\ell \cot\left( \Theta_{\rm g} - \frac{\pi}{2} - \mathrm{i}\mu_{\rm g}\right)\right),
    \label{eq: Phi_g}
\end{gather}
where $\mathrm{i}$ is the imaginary unit, $q_\ell$ is the coupling strength of the modes of degree $\ell$ \citep[e.g.,][]{shibahashi79,takata16a}, $\mu_{\rm g}$ is the amplitude modification that measures the efficiency of the damping process that operates in the g-mode cavity (see Eq. \ref{eq: mu_g}). The phase corresponding to the g-modes is given by the expression \citep[e.g.,][]{mosser+12b}
\begin{gather}
    \Theta_{\rm g} = \frac{\pi}{\Delta\Pi_\ell \ \nu},
    \label{eq: theta g}
\end{gather}
where $\Delta\Pi_\ell$ is the asymptotic period spacing of the g-modes.
The hight of the peaks is normalized due to the division by the peak hight of pure p-modes, $\mathcal{H}_{\rm p}$, which is given by
\begin{gather}
    \mathcal{H}_{\rm p} = \frac{2}{\left(1 - e^{-2\mu_{\rm p}} \right)^2}.
    \label{eq: peak height}
\end{gather}
In this expression, $\mu_{\rm p}$ is the amplitude modification that measures the efficiency of the damping process that takes place in the p-mode cavity (see Eq. \ref{eq: mu_p}). Therefore, all resolved oscillation modes that are not damped in the g-mode cavity appear as peaks with height one in the normalized power spectrum, while the modes that are damped in the g-mode cavity have a height of less than one. 
The remaining parameter $A$ in Eq. \eqref{eq: normalized power spectrum} is the internal resonance enhancement \citep{muller+26}, which measures how interference affects the amplitude of the waves in the system as a function of the oscillation frequency:
\begin{gather}
    A = \left|1 - e^{\mathrm{i}2\left(\Theta_{\rm p} + i\mu_{\rm p} + \Phi_{\rm g}\right)}\right|^{-2}.
    \label{eq: internal resonace enhancement factor}
\end{gather}
The phase corresponding to the p-modes can be expressed as \citep[e.g.,][]{mosser+12b}
\begin{gather}
    \Theta_{\rm p} = \pi \left( \frac{\nu}{\Delta\nu} - \frac{\ell}{2} + d_{0\ell} \right).
    \label{eq: theta_p}
\end{gather}
Here, $\Delta\nu$ is the large frequency separation and $d_{0\ell}$ is a non-radial correction, which we have selected as $d_{00}=0$, $d_{01}=-0.02$, and $d_{02} = 0.12$, in agreement with the values chosen by \citet{stello+16a,stello+16b} \citep[see also][]{huber+09}.
In Eqs.~\ref{eq: theta g} and \ref{eq: theta_p}, we neglected the phase shifts $\epsilon_{\rm g}$ and $\epsilon_{\rm p}$ for simplicity. This does not strongly affect the general trends in the mode visibilities \citep[see][]{muller+26}.

Since the power excess of solar-like oscillators is not uniform but bell-shaped and centered around the frequency of maximum oscillation power, $\nu_{\rm max} $, we follow \citet{muller+26} and multiply $P_\ell$ by a Gaussian function, $G$, centered around $\nu_{\rm max}$ with height one and standard deviation $\sigma_{\rm env}$ to obtain the relative observable power spectrum\footnote{The relative observable power spectrum $\mathcal{P}_\ell$ defined here differs from the relative power spectrum $\mathcal{P}$ in \citet{muller+26} by an additional factor $\mathcal{S}_{\ell}$.}:
\begin{gather}
    \mathcal{P}_{\ell} = \mathcal{S}_{\ell}\, G\, P_{\ell}.
\end{gather}
The standard deviation of the Gaussian distribution, $\sigma_{\rm env}$, is estimated using the empirical scaling relation from \citet{mosser+12}, which depends on the stellar mass, $M_\star$, and $\Delta\nu$.
By definition, the spatial response is $\mathcal{S}_{\ell=0} = 1$ for the radial modes, and for the dipole and quadrupole modes, \citet{ballot+11} have calculated theoretical values $\mathcal{S}_{\ell}$ that are expected for observations with {\it Kepler}.
To obtain the complete PSD, in which modes of all $\ell$ are present, we sum the contributions of the $\ell$ considered in this work, which reads
\begin{gather}
    \mathcal{P}_{\rm tot} = \mathcal{P}_{\ell=0} + \mathcal{P}_{\ell=1} + \mathcal{P}_{\ell=2}.
\end{gather}
The contributions corresponding to the oscillations of degree $\ell \geq 3$ are negligible for the analysis carried out in this work, since their observable amplitudes are very small compared to the lower-degree modes \citep[e.g.,][]{ballot+11,mosser+12}. Therefore, they are not taken into account.

\subsection{Visibility} \label{sect: visibility}

\begin{table}[]
    \caption{Segments in the PSD dominated by modes of different spherical degrees.}
    \centering
    \begin{tabular}{ccc}
    \hline\hline
    $\ell$ & $\nu_\ell^{\rm low}$ & $\nu_\ell^{\rm high}$ \\
    \hline
    0 & $\nu_0 - 0.06\Delta\nu$ & $\nu_0 + 0.10\Delta\nu$ \\
    1 & $\nu_0 + 0.25\Delta\nu$ & $\nu_0 + 0.78\Delta\nu$ \\
    2 & $\nu_0 - 0.22\Delta\nu$ & $\nu_0 - 0.06\Delta\nu$ \\
    3 & $\nu_0 + 0.10\Delta\nu$ & $\nu_0 + 0.25\Delta\nu$ \\
    \hline
    \end{tabular}
    \tablefoot{The parameter $\nu_0$ denotes the frequency of a radial mode. The lower and upper boundaries of the segments $\nu_\ell^{\rm low}$ and $\nu_\ell^{\rm high}$ were adopted from \citet{stello+16a,stello+16b}. In this work, we do not consider octopolar modes (i.e., modes with $\ell=3$), since their amplitudes are very small compared to the lower-degree modes \citep[e.g.,][]{ballot+11,mosser+12}.}
    \label{tab: l segments}
\end{table}

The visibility can be calculated directly from the PSD as the ratio of the integrated power spectrum of the multipole modes to that of the radial modes over the frequency range under consideration, which in this work ranges from $\nu_{\rm left} = \nu_{\rm max}-2\Delta\nu$ to $\nu_{\rm right}=\nu_{\rm max}+2\Delta\nu$. The values of $\nu_{\rm left}$ and $\nu_{\rm right}$ were chosen to match the frequency range considered in the observational studies by \citet{stello+16a,stello+16b}. 

\subsubsection{Theoretical visibility}

From a theoretical point of view, we can directly use the normalized power spectrum to obtain an estimate for visibility:
\begin{gather}
    V^2_\ell = \frac{ \int_{\nu_{\rm left}}^{\nu_{\rm right}} P_\ell \,\mathrm{d}\nu }{ \int_{\nu_{\rm left}}^{\nu_{\rm right}} P_{\ell=0} \,\mathrm{d}\nu }.
    \label{eq: theoretical visibility}
\end{gather}
However, two assumptions are made when calculating the visibility according to Eq. \eqref{eq: theoretical visibility}. First, $V^2_\ell$ does not take into account $\mathcal{S}_{\ell}$, which means that the maximum possible value of $V^2_\ell$ is approximately one. Second, the bell-shape of the power excess of the oscillations is not taken into account\footnote{The visibility defined in \citet{muller+26} takes into account the influence of the bell-shape of the power excess, except in their Appendix D, where the definition of the visibility is similar to that in Eq. \eqref{eq: theoretical visibility}.}.

\subsubsection{Observable visibility}

Observed PSDs of real stars exhibit modes with multiple $\ell$ whose contributions cannot be completely separated from each other. Due to this, the theoretical visibility, $V^2_\ell$, cannot be determined from real data.
In observations, it is thus common practice to divide the range of observable frequencies into segments in which modes of a specific $\ell$ are expected to dominate in order to isolate the modes of a specific $\ell$ and minimize contamination effects due to modes of different degrees \citep[e.g.,] []{mosser+12, stello+16a, stello+16b, mosser+17, dreau+21}. The visibility is then estimated by integrating the power spectrum over these segments rather than over the entire frequency range. 
This can lead to the above-mentioned bias, where mixed modes of a given $\ell$ are not taken into account in the calculation of the visibility corresponding to that $\ell$ and instead contribute to the visibility of another spherical degree.

Applying the same method to our synthetic PSDs, the observable visibility, $\mathcal{V}^2_\ell$, can be calculated by integrating the total relative power spectrum over the segments of the frequency range:
\begin{gather}
    \mathcal{V}^2_\ell = \frac{\sum\limits_{\nu_i \in \{\nu_0\}} \left( \int_{\nu_i+\nu_\ell^{\rm low}}^{\nu_i+\nu_\ell^{\rm high}} \mathcal{P}_{\rm tot} \,\mathrm{d}\nu \right)}{\sum\limits_{\nu_i \in \{\nu_0\}} \left( \int_{\nu_i+\nu_{\ell = 0}^{\rm low}}^{\nu_i+\nu_{\ell = 0}^{\rm up}} \mathcal{P}_{\rm tot} \,\mathrm{d}\nu \right)}.
\end{gather}
Here, $\nu_\ell^{\rm low}$ ($\nu_\ell^{\rm high}$) is the lower (upper) boundary of the segment of the frequency range associated with the degree $\ell$, and $\{\nu_0\}$ is the set of radial mode frequencies. We only account for the contribution of the segments that fall within the frequency range under consideration (i.e., $\nu_{\rm left} \leq \nu \leq \nu_{\rm right}$).
The choice of boundaries $\nu_\ell^{\rm low}$ and $\nu_\ell^{\rm high}$ is somewhat arbitrary, and different definitions have been used in the literature. In this work, we adopt the boundaries used by \citet{stello+16a,stello+16b}, listed in Table \ref{tab: l segments}.

\subsection{Stellar models} \label{sect: stellar models}

We used version r23.05.1 of the publicly available stellar evolution code MESA \citep[Modules for Experiments in Stellar Evolution;][]{mesa1,mesa2,mesa3,mesa4,mesa5,mesa6} to calculate three evolutionary tracks with stellar masses of $M_\star = 1$, 1.25, and $1.5\ M_\odot$. The metallicity was set to $Z_{\rm init} = 0.02$. The inputs to our stellar model are deliberately kept simple, as we are only interested in general trends and are not attempting to model individual stars in detail. We include exponentially decaying overshoot below the convective envelope using the input parameters \texttt{overshoot\_f=0.015}, \texttt{overshoot\_f0=0.005} \citep[e.g.,][]{khan+18}, and \texttt{overshoot\_D\_min=1d-2} \citep{buchele25}, which smooths the stellar structure. Smooth structure profiles are crucial for calculating the coupling strength, $q_\ell$ \citep[e.g.,][]{takata16a,vanRossem+24,vanLier+25}, which is described in Appendix \ref{app: coupling}. 
In the figures displayed in the following, we show the evolution of the stars on the RGB using 59 models whose frequency of maximum oscillation power is approximately 5 $\mu{\rm Hz}$ apart and lies in the range $10\ \mu{\rm Hz} \leq \nu_{\rm max} \leq 300\ \mu{\rm Hz}$. The uniform resolution in $\nu_{\rm max}$ was chosen because we present the evolution on the RGB as a function of $\nu_{\rm max}$ in this work. This means, in particular, that we do not resolve the luminosity bump on the RGB in our figures (even though it was modeled). However, even if the bump is shown, the predicted visibility for a given value of $\nu_{\rm max}$ does not change strongly \citep[e.g.,][]{cantiello+16}, which is why we refrain from showing it explicitly.

\subsection{Damping processes} \label{sect: damping processes}

Oscillations in red giant stars are intrinsically damped. The efficiency of the damping processes is measured using the amplitude modifications $\mu_{\rm p}$ and $\mu_{\rm g}$, which were introduced in Sect. \ref{sect: power spectrum}. These parameters are directly related to the damping rates of the processes occurring in the p-mode cavity, $\eta_{\rm p}$, and in the g-mode cavity, $\eta_{\rm g}$, via \citep{muller+26}
\begin{gather}
    \mu_{\rm p} = \frac{\eta_{\rm p}}{2\ \Delta\nu}
    \label{eq: mu_p}
\end{gather}
and
\begin{gather}
    \mu_{\rm g} = \frac{\eta_{\rm g}}{2\ \Delta\Pi_\ell\ \nu^2}.
    \label{eq: mu_g}
\end{gather}
In the following subsections, we determine the amplitude modification corresponding to each process considered in this work.

\subsubsection{Damping by the interaction with convection} \label{sect: convective damping}

\begin{table}[]
    \caption{Exponents for the scaling relationships of the radial mode linewidth.}
    \centering
    \begin{tabular}{cccc}
    \hline\hline
    $M_\star\ /\ M_\odot$ & 1 & 1.25 & 1.5 \\
    \hline
    $\alpha_T$ & 4.55 & 5.29 & 4.54 \\
    $\alpha_M$ & 0.251 & 0.251 & 0.251 \\
    \hline
    \end{tabular}
    \tablefoot{The scaling relation used for the radial mode linewidth is given by $\Gamma_0 \propto (T_{\rm eff}/ {\rm K})^{\alpha_T } (M_\star / M_\odot)^{\alpha_M}$. The exponents listed in the table were determined by \citet{vrard+18}. Note that we neglect the $T_{\rm eff}$-dependence of $\alpha_M$.}
    \label{tab: exponents gamma0}
\end{table}
\begin{figure}[]
    \centering
    \resizebox{\hsize}{!}{\includegraphics{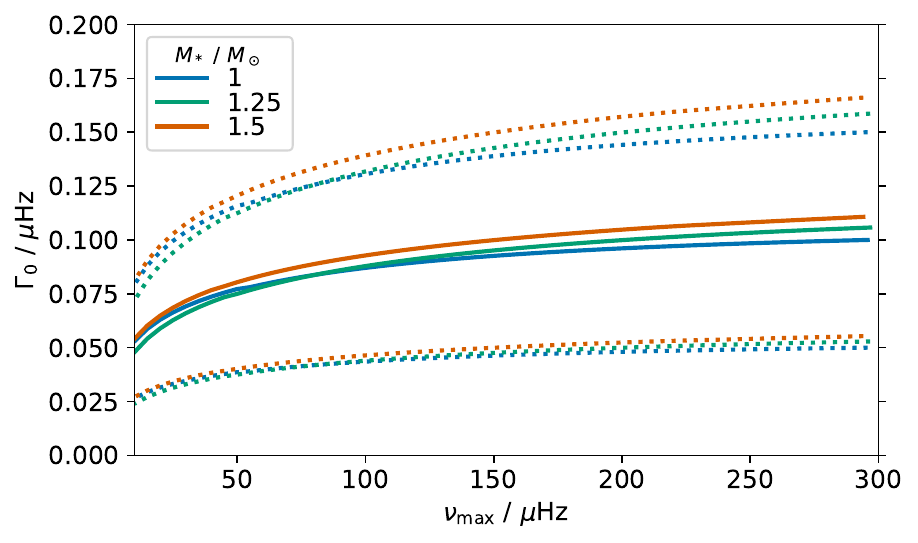}}
    \caption{Radial mode linewidth as a function of the frequency of maximum oscillation power. Colors indicate different stellar masses. For each mass, we use three reference values for the linewidth so that the tested values approximately encompass the range of observed linewidths. The intermediate value is represented by the solid line, the upper and lower values by the dotted lines.}
    \label{fig: evolution linewidth}
\end{figure}

In solar-like oscillators, the oscillation modes are both driven and damped by interaction with convection in the upper part of the convective envelope (i.e., $\mu_{\rm p} > 0$ for all RGB stars). In this region, the oscillation periods are comparable to the timescales of the most energetic convective eddies, which means that the fluctuations caused by the oscillations have a feedback effect on the properties of the turbulence, such as convective flux and Reynolds stress \citep[e.g.,][]{dupret+09}. A proper treatment of the interaction of oscillations with convection is quite complex and requires non-adiabatic numerical calculations \citep[e.g.,][]{unno67,gough77,gabriel96,xiong+97,grigahcene05}, which are beyond the scope of this work.

Instead of calculating the damping rates numerically, we use observational constraints to determine $\mu_{\rm p}$. Damped oscillations appear in the PSD as Lorentzian functions with a specific width, which is usually measured as the full width at half maximum or mode linewidth, $\Gamma$. The linewidth is a direct measure of the damping rate of an oscillation mode. Since the radial modes are pure p-modes, we can relate their linewidth, $\Gamma_0$, directly to the energy loss caused by the interaction of the oscillations with convection via $\eta_{\rm p} = \pi\Gamma_0$ \citep[][]{samadi+15}. Assuming that $\eta_{\rm p}$ does not depend on $\ell$, the amplitude modification in the p-mode cavity is given by \citep{takata16b, muller+26}
\begin{gather}
    \mu_{\rm p} = \frac{\pi\ \Gamma_0}{2 \ \Delta\nu}.
    \label{eq: mu_p convective}
\end{gather}

The radial mode linewidth has been measured for many solar-like oscillators \citep[e.g.,][]{baudin+11,corsaro+12,corsaro+15,vrard+18}. From a theoretical point of view, the simplest approach is to model $\Gamma_0$ as one or more constant values that roughly match the observed range of $\Gamma_0$.
Here, we follow a slightly more sophisticated approach and scale $\Gamma_0$ as a function of the effective temperature, $T_{\rm eff}$, according to $\Gamma_0 \propto (T_{\rm eff}/ {\rm K})^{\alpha_T}$ and as a function of the stellar mass according to $\Gamma_0 \propto (M_\star / M_\odot)^{\alpha_M}$. We adopt the exponents $\alpha_T$ and $\alpha_M$ as determined by \citet{vrard+18} using their PROB method and for simplicity we neglect the uncertainties of these exponents. As a reference point, we use the value of $\Gamma_0$ from the model with a mass of $M_\star = 1\ M_\odot$ and a frequency of maximum oscillation power $\nu_{\rm max} \approx 300\ \mu$Hz.
To account for the range of possible values for $\Gamma_0$, which can differ even for stars with similar values for $T_{\rm eff}$ and $M_\star$, we select three reference values ($\Gamma_0 = 0.05$, $0.1$ and $0.15\ \mu{\rm Hz}$), which are then scaled using the proportionalities. The exponent $\alpha_T$ depends on the stellar mass and is different for each evolutionary track (see Table \ref{tab: exponents gamma0}). For the exponent $\alpha_M$, we use a fixed value of 0.251, which was estimated by \citet{vrard+18} for stars with an effective temperature between 4700 and 4900 K. We do not consider other values for $\alpha_M$ in order to ensure a smooth behavior of $\Gamma_0$ as the star ascends the RGB. The resulting estimates of $\Gamma_0$ are shown in Fig. \ref{fig: evolution linewidth}.
In this approach, we neglect the variation of $\Gamma_0$ with $\nu$. However, when only considering the four most central pressure radial orders, as we do here, this dependence is rather weak \citep[e.g.,][]{vrard+18} and should not significantly affect our results.

\subsubsection{Radiative damping (rad)}

\begin{figure}[]
    \centering
    \resizebox{\hsize}{!}{\includegraphics{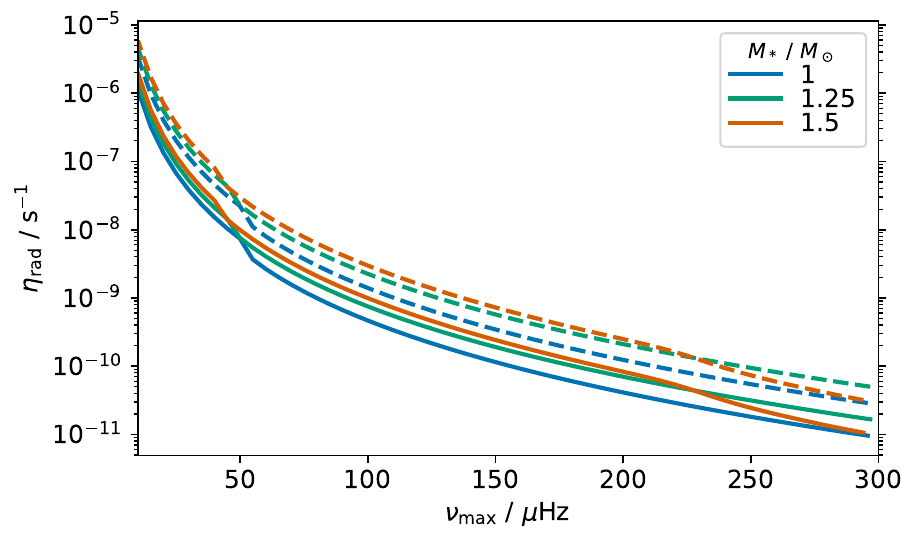}}
    \caption{Damping rate corresponding to radiative damping of the dipole and quadrupole modes as a function of the frequency of maximum oscillation power. Colors indicate different stellar masses. For each mass, the damping rate of the dipole modes is represented by the solid line, the damping rate of the quadrupole modes by the dashed line.}
    \label{fig: evolution radiative damping}
\end{figure}

In the remainder of Sect. \ref{sect: damping processes}, we focus on damping processes that occur in the g-mode cavity and therefore contribute to $\mu_{\rm g}$. One damping process that is expected to occur in all RGB stars is heat loss due to fluctuations in the temperature gradient caused by oscillations in the radiative core (i.e., “radiative damping”). Under asymptotic conditions, the damping rate of radiative damping is given by \citep[e.g.,][]{dziembowski77,vanHoolst+98,dziembowski+01,hekker+17}
\begin{gather}
    \eta_{\rm rad} = \frac{(\ell(\ell+1))^{3/2}}{64\ \pi^4\nu^3} \frac{1}{\int_0^{r_{\rm core}} k_r \,\mathrm{d}r} \int_0^{r_{\rm core}} \frac{\nabla_{\rm ad} (\nabla_{\rm ad} - \nabla)}{\nabla} \frac{NgL}{pr^5} \,\mathrm{d}r,
\end{gather}
where $r$ is the radial coordinate, $r_{\rm core}$ is the base of the convective outer layer, $N$ is the Brunt–Väisälä frequency, $g$ is the local gravitational acceleration, $L$ is the local luminosity, and $p$ is the pressure. The real and the adiabatic temperature gradients are denoted by $\nabla$ and $\nabla_{\rm ad}$, respectively. The resulting values of $\eta_{\rm rad}$ are shown in Fig. \ref{fig: evolution radiative damping}.
Radiative damping is generally negligible for modes with low $\ell$ in early RGB stars, but becomes more significant as the star evolves due to the increasing density contrast between the core and the envelope. Using Eq. \eqref{eq: mu_g}, the damping rate can be converted into a corresponding amplitude modification
\begin{gather}
    \mu_{\rm g,rad} \approx \frac{(\ell(\ell+1))^{3/2}}{128\ \pi^5\nu^4} \int_0^{r_{\rm core}} \frac{\nabla_{\rm ad} (\nabla_{\rm ad} - \nabla)}{\nabla} \frac{NgL}{pr^5} \,\mathrm{d}r,
\end{gather}
where we have approximated $\int_0^{r_{\rm core}} k_r \,\mathrm{d}r \approx \Theta_{\rm g}$\footnote{This is based on the assumption $\omega^2 \ll S_\ell^2, N^2$ and leads to deviations from the original value of less than 2\% on the early RGB, where the effect of radiative damping is negligible. The deviations have no significant impact on our results, as they decrease with decreasing $\nu_{\rm max}$.} (see Eq. \ref{eq: theta g}).

\subsubsection{Magnetic suppression of mixed modes (SUP)}

In the following subsections, we present three prescriptions describing the damping process caused by the interaction of internal gravity waves with a strong magnetic field in the core of RGB stars. To distinguish them from radiative damping (rad), we use capitalized abbreviations for the magnetic damping (e.g., SUP).
\citet{fuller+15} suggest that a core magnetic field could constitute a strong damping mechanism once the magnetic field strength exceeds a threshold value. They define this so-called critical field strength as
\begin{gather}
    B_{\rm crit} = \sqrt{\pi\rho} \frac{r}{N} \frac{4 \pi^2 \nu^2}{\sqrt{\ell(\ell+1)}}.
    \label{eq: B crit}
\end{gather}
The amplitude modification (and the damping rate) corresponding to this magnetic suppression of mixed modes is given by a step-function:
\begin{gather}
    \mu_{\rm g,SUP} = 
    \begin{cases}
     0, & \text{if}\ B < B_{\rm crit},\\
     \infty, & \text{if}\ B \geq B_{\rm crit}.
    \end{cases}
    \label{eq: damping rate SUP}
\end{gather}
Here, $B$ is the strength of the internal magnetic field. Since $B_{\rm crit}$ depends on $\nu$ and $\ell$, $\mu_{\rm g,SUP}$ varies over the PSD of a star with a magnetic field of a given strength $B$.

If $\mu_{\rm g,SUP} = \infty$, all of the mode energy that leaks from the p- into the g-mode cavity is lost. This complete suppression of mixed modes in the g-mode cavity has been predicted in several theoretical works \citep{fuller+15,lecoanet+17,rui+fuller23,david+26}, and the inferred visibilities roughly agree with observations \citep{stello+16a,stello+16b}. However, \citet{mosser+17} show that a significant fraction of depressed dipole modes stars exhibit dipolar mixed modes. This means that some of the mode energy must be able to escape from the g-mode cavity and contribute to the observable oscillations, which in turn means that $\mu_{\rm g}$ is certainly finite in these stars. To account for these observational constraints, we introduce a modified version of magnetic suppression in the following subsection, which allows for finite non-zero damping rates.

\subsubsection{Ad hoc depression of mixed modes (DEP)} \label{sect: modified depression of mode energy}

To allow for an observable signature of the mixed mode in depressed dipole mode stars, we test an ad hoc modification of the previously discussed scenario of complete suppression. In particular, we set a finite upper limit on the amplitude modification by requiring that 25\% of the mode energy entering the g-mode cavity is allowed to return to the p-mode cavity, while 75\% of the mode energy entering the g-mode cavity is dissipated. The choice of these percentages is arbitrary due to the lack of theoretical constraints, so that we consider the scenario described in this section to be a qualitative test of concept. One possible explanation for the presence of mixed modes in depressed dipole mode stars is that the interaction of the waves with the magnetic field inherently allows a small fraction of the mode energy to return from the g-mode cavity (see also Sect. \ref{sect: ray tracing} on Hamiltonian ray tracing, which predicts this behavior).
Another possible scenario is that the energy loss caused by interaction with the magnetic field is complete, but a small portion of the mode energy is reflected back into the p-mode cavity before being dissipated by the magnetic field. However, depending on the location in the g-mode cavity where the reflection occurs, this would lead to a shift in $\Delta\Pi_\ell$ (see Sect. \ref{sect: location of magnetic damping}), for which there is currently no concrete observational evidence \citep{mosser+17}.

The fraction of mode energy retained by the wave after an energy dissipation event in the g-mode cavity can be expressed as the squared modulus of the reflection coefficient at the lower boundary of the g-mode cavity ($|R_{\rm g, DEP}|^2 = 0.25$, which yields $|R_{\rm g,DEP}| = 0.5$). The reflection coefficient is directly related to the amplitude modification via \citep{takata16b,muller+26}
\begin{gather}
    \max\left(\mu_{\rm g,DEP}\right)= - \frac{1}{2} \ln(|R_{\rm g,DEP}|) \approx 0.347,
\end{gather}
such that
\begin{gather}
    \mu_{\rm g,DEP} = 
    \begin{cases}
     0, & \text{if}\ B < B_{\rm crit},\\
     0.347, & \text{if}\ B \geq B_{\rm crit}.
    \end{cases}
\end{gather}

\subsubsection{Ray tracing of magneto-gravity waves (RT)} \label{sect: ray tracing}

When performing a Hamilton ray tracing analysis of magneto-gravity waves in the radiative core of RGB stars \citep{loi20a,loi20b,muller+25}, a gradual transition from no energy loss to very high energy loss is obtained as a function of $B$.
Although this gradual increase in the efficiency of the damping mechanism is exactly what is required to allow the existence of mixed modes in the PSDs of depressed dipole modes stars, the Hamilton ray tracing approach relies on the assumption that the horizontal wavenumber of the oscillations has a lengthscale much smaller than that of the horizontal variation of the magnetic field. This assumption is not appropriate when studying observable modes with low $\ell$. However, the partial dissipation of mode energy prevails when studying waves with much higher $\ell$ that have smaller horizontal lengthscales \citep{muller+25}, for which the assumptions made in the ray tracing formalism are expected to be suitable.
In addition, the ray tracing formalism takes into account the inner turning point of the g-mode cavity, in contrast to the aforementioned works predicting complete suppression in the g-mode cavity. We discuss the importance of the inner turning point in Sect. \ref{sect: location of magnetic damping}.

In the first article of this series of publications \citep{muller+25}, we made the predictions of the Hamilton ray tracing approach developed by \citet{loi20a} testable on a larger scale by fitting an empirical function that describes the fraction of mode energy lost due to the magnetic field (i.e., the trapped fraction $f_{\rm T}$) as a function of magnetic field strength \citep{muller+25}:
\begin{gather}
    f_{\rm T}(B) = \max \left[ \ 0,\ 1 - 0.635\ \left( \frac{B}{B_{\rm crit,RT}} \right)^{-1} \right. \notag\\
    \qquad\qquad\qquad \left. +\, 0.438\ \left( \frac{B}{B_{\rm crit,RT}} \right)^{-2} - 0.272\ \left( \frac{B}{B_{\rm crit,RT}} \right)^{-3} \right].
\end{gather}
In this expression, we use the critical field strength as defined by \citet{loi20a}, which is given by
\begin{gather}
    B_{\rm crit,RT} = 2\ B_{\rm crit}.
\end{gather}
The portion of energy that is dissipated can be expressed by using a reflection coefficient $|R_{\rm g,RT}|$ via $f_{\rm T} = 1 - |R_{\rm g,RT}|^2$ (see Sect. \ref{sect: modified depression of mode energy}). The corresponding amplitude modification reads
\begin{gather}
    \mu_{\rm g,RT} = - \frac{1}{2} \ln(|R_{\rm g,RT}|) = - \frac{1}{4} \ln(1 - f_{\rm T}).
    \label{eq: mu_g from f_T}
\end{gather}

\subsubsection{Location of the magnetic damping} \label{sect: location of magnetic damping}

\begin{figure}[]
    \centering
    \resizebox{\hsize}{!}{\includegraphics{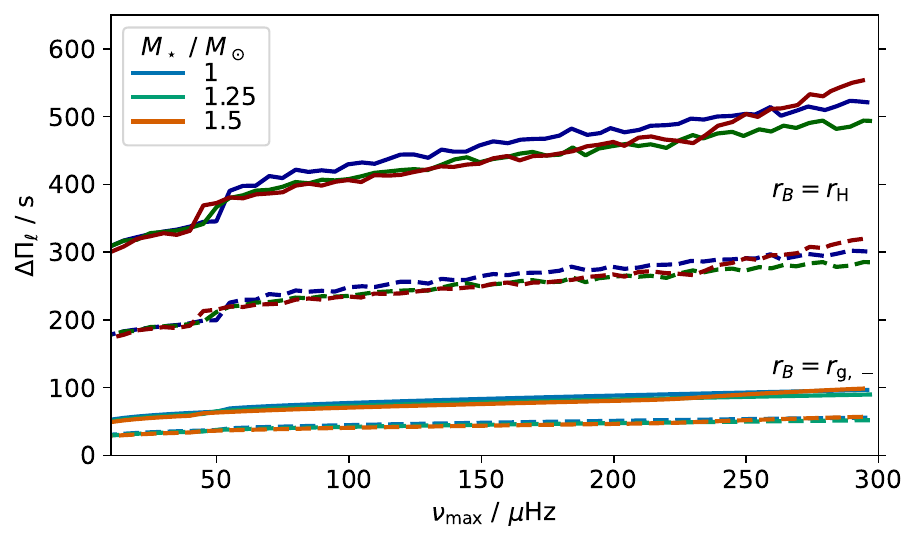}}
    \caption{Asymptotic period spacing as a function of the frequency of maximum oscillation power for two effective sizes of the g-mode cavity. The solid (dashed) lines correspond to the dipole (quadrupole) modes. The colors represent different stellar masses. Darker colors indicate that the hydrogen-burning shell was used as the lower boundary of the g-mode cavity.}
    \label{fig: period spacing rB}
\end{figure}

The three prescriptions for the magnetic damping discussed in the previous subsections depend on the ratio of the critical field strength and the actual field strength. Both $B_{\rm crit}$ and $B$ are functions of the radius. Therefore, we must select the location at which we expect $B$ to first approach $B_{\rm crit}$ and compare the two field strengths at the corresponding radius, $r_B$. There are two natural choices for $r_B$: the hydrogen-burning shell, at which $B_{\rm crit}$ has a local minimum \citep[$r_B = r_{\rm H}$; e.g.,][]{fuller+15}, and the inner turning point of the g-mode cavity, which is close to the stellar center where most magnetic field configurations (e.g., a dipolar field) reach their maximum value ($r_B = r_{\rm g,-}$). 
Assuming that the downward-traveling waves are either reflected upward or dissipated when reaching $r_B$, this radius functions as an effective inner turning point of the oscillations in the g-mode cavity, thus modifying the size of the cavity. Therefore, the value of the observable asymptotic period spacing depends on the choice of $r_B$.

In Fig. \ref{fig: period spacing rB}, we show the value of the asymptotic period spacing for the two options of $r_B$. While we recover the nominal value for $r_B = r_{\rm g,-}$, the period spacing is an order of magnitude larger if $r_B = r_{\rm H}$. \citet{mosser+17} show that the observed period spacings of the depressed dipole mode stars are comparable to those of the normal stars. In fact, $\Delta\Pi_{\ell=1} < 100$ s for all RGB stars with depressed mixed modes in their sample, which indicates that $r_{\rm g,-}$ is the appropriate choice for $r_{B}$. However, the observed value of $\Delta\Pi_\ell$ can also be modified directly by the magnetic field, as it constitutes an additional restoring force to the oscillations \citep[e.g.,][]{loi20b,bugnet22}. Observations by \citet{deheuvels+23,deheuvels+26} show that magnetic fields strong enough to cause mode depression can decrease the observed value of $\Delta\Pi_{\ell=1}$ by less than a few tens of seconds. Although even stronger internal magnetic fields could lead to a greater decrease in the observed $\Delta\Pi_{\ell=1}$, it is questionable whether this could account for a difference of several hundred seconds, which would be necessary to compensate for the increase in $\Delta\Pi_{\ell=1}$ caused by setting $r_B = r_{\rm H}$.
Furthermore, \citet{deheuvels+26} report the first observational constraints on the radial extent of the internal magnetic field in red giant stars. They find that the field is well confined to the region inside the hydrogen-burning shell and should not have a significant strength at the hydrogen-burning shell.

For these reasons, we assume that the magnetic field approaches $B_{\rm crit}$ at the stellar center and set $r_B = r_{\rm g,-}$ in this study. This is in line with the behavior of the so-called Prendergast field, which seems to first approach $B_{\rm crit}$ at the center in models of RGB stars \citep[][]{muller+25}. Since we parameterize the magnetic field using a measure close to the central field strength, the values of $B$ reported in this work are higher than the minimum weighted averages inferred from observations \citep[e.g.,][]{li+22,li+23,hatt+24}, which are most sensitive to the hydrogen-burning shell. \citet{deheuvels+26} observationally constrained the central field strengths of eight red giant stars and found that they are much higher than the corresponding weighted averages, reaching values up to several MG ("Mega-Gauss").

Another important aspect is that the theoretical studies predicting complete dissipation of mode energy upon interaction with a strong magnetic field \citep[i.e.,][]{lecoanet+17,rui+fuller23,david+26} assume that the magnetic field strength becomes critical deep inside the g-mode cavity. In other words, these studies neglect the presence of the inner turning point of the g-mode cavity\footnote{In the Cowling approximation, the inner turning point of the g-mode cavity is characterized by the fact that the angular oscillation frequency, $\omega$, is equal to the local buoyancy frequency, $N(r_{\rm g,-})$. This implies that the radial wavenumber is $k_r=0$ at the inner turning point, whereas the horizontal wavenumber, $k_{\rm h}$, remains larger than zero. \citet{rui+fuller23} assume a hierarchy of variables, which includes $\omega \ll N$ and $k_r \gg k_{\rm h}$. Clearly, this cannot be satisfied in the vicinity of the turning point. \citet{lecoanet+17} and \citet{david+26} assume that $N$ is either a constant or increasing function with depth, which means that the angular frequency of a downward-traveling wave does not approach the local $N$. They also added a layer of artificial damping to the bottom of their simulation box, so that downward-traveling waves that reach this layer are dissipated instead of being reflected.}.
While this assumption is likely justified if $r_B = r_{\rm H}$, it might be inappropriate when $B$ approaches $B_{\rm crit}$ close to the turning point (i.e., $r_B = r_{\rm g,-}$).

In this context, it is noteworthy that the studies using the ray tracing method, which accounts for the presence of the inner turning point, predict partial dissipation of mode energy \citep[][]{loi20a,muller+25}. This suggests that accounting for the inner turning point of the g-mode cavity might be essential to explain the existence of the depressed dipole mode stars with an observable mixed mode signature. 
The data from \citet{muller+25} support this interpretation: the reflected rays are those that encounter the inner turning point of the g-mode cavity where $k_r=0$, while the non-reflected rays are trapped at a radius larger than that of the turning point and thus never reach it. The assumptions underlying the ray tracing analysis are expected to be appropriate for modes with high $\ell$. It therefore seems reasonable that the inner turning point of the g-mode cavity should also be taken into account when investigating the behavior of the observable modes with low $\ell$.

\subsection{Initialization} \label{sect: initialization}

To calculate the power spectrum and visibility as described in Sects. \ref{sect: power spectrum} and \ref{sect: visibility}, we need to constrain the following stellar parameters: $M_\star$, $T_{\rm eff}$, $\nu_{\rm max}$, $\Delta\nu$, $\Delta\Pi_{\ell}$, $q_\ell$, $\mu_{\rm p}$, and $\mu_{\rm g}$. The two amplitude modifications can essentially be attributed to $\Gamma_0$, $\eta_{\rm rad}$, $B$, and $B_{\rm crit}$.
All of these parameters, with the exception of $\Gamma_0$ and $B$, can be calculated directly from our stellar models.
We emphasize that although we use numerical stellar models, we do not employ an oscillation code to calculate the eigenfrequencies of the oscillation modes. Even though non-adiabatic calculations are generally possible, current oscillation codes cannot account for the dissipation of mode energy caused by a strong internal magnetic field. Instead, we use the asymptotic procedure described in Sect. \ref{sect: power spectrum}, which allows us to compute synthetic PSDs directly from the damping rates discussed in Sect. \ref{sect: damping processes}.

The coupling strength, $q_\ell$, requires special care. \citet{shibahashi79} and \citet{takata16a} have each derived an asymptotic approximation for $q_{\ell=1}$ that is valid when the evanescent zone is either wide or very narrow, respectively. However, no asymptotic expression is known for the case of an evanescent zone with intermediate width. When a star ascends the RGB, it is expected to undergo all three regimes, transitioning from strong coupling and a very thin evanescent zone to weak coupling and a wide evanescent zone. 
\citet{vanLier+25} report that the weak coupling prescription proposed by \citet{shibahashi79} approximates the coupling strength of stars with an evanescent zone of intermediate width with reasonable accuracy. However, the prescriptions of \citet{shibahashi79} and \citet{takata16a} generally do not smoothly transition into each another during the evolution of the star \citep{pincon+20,jiang+22,vanLier+25}, which would lead to an abrupt change in $q_{\ell=1}$. Since such a discontinuity in $q_{\ell=1}$ has not been observed in real stars \citep[e.g.,][]{mosser+17coupling} and would lead to artifacts in our predictions for the visibilities, a direct transition between the two prescriptions is inappropriate in the context of this work. Instead, we use a linear interpolation to connect the two prescriptions in the range of an evanescent zone with an intermediate width. The calculation of $q_\ell$ is described in detail in Appendix \ref{app: coupling}. Our inferred vales of $q_\ell$ are shown in Fig. \ref{fig: evolution coupling} and are consistent with estimates of previous works \citep[][]{vanLier+25}.

The two remaining input parameters are the radial mode linewidth, $\Gamma_0$, and the magnetic field strength, $B$. The calculation of $\Gamma_0$ is described in Sect. \ref{sect: convective damping}.
For the magnetic field strength $B$, we consider a wide range of initial values: $B_{\rm MS} = 10$, 18, 32, 56, 100, 180, 320, 560, and 1000 kG. These values correspond to the field strength at the end of the main sequence. The magnetic field strength is scaled along the evolution on the RGB using magnetic flux conservation, whereby $B$ increases with decreasing $\nu_{\rm max}$. As discussed in Sect. \ref{sect: location of magnetic damping}, we always measure $B$ and $B_{\rm crit}$ at the inner turning point of the g-mode cavity (i.e., close to the stellar center).

\section{Spatial response} \label{SECT: spatial response}

Before we can calculate the visibilities, we must first constrain the spatial response, $\mathcal{S}_{\ell}$ (see Sect. \ref{sect: power spectrum}). This factor indicates the expected theoretical visibility, assuming that the total energy of the modes is evenly distributed across all $\ell$ (i.e., energy equipartition).
For observations with {\it Kepler}, \citet{ballot+11} predicted that this factor should be approximately $\mathcal{S}_{\ell=1} = 1.54$ and $\mathcal{S}_{\ell=2} = 0.58$ for dipole and quadrupole modes, respectively (assuming $T_{\rm eff} = 4800$ K). However, in observations, the average dipole and quadrupole mode visibilities of normal stars (i.e., stars with normal multipole mode amplitudes) deviate from these predictions \citep[e.g.,][]{stello+16a,stello+16b}. While the amplitudes of the dipole modes are smaller than expected (1.35), the quadrupole modes appear to be larger (0.69). As mentioned before, \citet{mosser+12} suggested that the absence of dipole mode energy could be due to the spread of gravity-dominated mixed dipole modes away from the location of the nominal p-mode, causing them to be neglected in the calculation of the observable visibility, which would result in a bias compared to the theoretical value. Following this line of reasoning, these gravity-dominated dipole modes could instead contribute to $\mathcal{V}_{\ell=2}^2$, which could potentially explain their increase relative to the predicted values. In this section, we put this hypothesis to the test.

\subsection{Method for estimating $\mathcal{S}_{\ell}$} \label{sect: method for estimating Gell}

To estimate the spatial response, we compare our predictions of the observable visibility, $\mathcal{V}_\ell^2$, with the actual observed visibilities from \citet{stello+16a,stello+16b}, which we denote as $(\mathcal{V}_\ell^2)^{\rm Stello}$.
Since both $\mathcal{V}_\ell^2$ and $(\mathcal{V}_\ell^2)^{\rm Stello}$ are subject to the above-mentioned bias, they can be compared directly.
As a first step, we calculated the normalized power spectrum for the degrees $\ell=0$, 1, and 2 for each RGB model on our evolutionary tracks and multiplied it by the Gaussian envelope. We did this for each of our three estimates for $\Gamma_0$, resulting in three values of the amplitude modification in the p-mode cavity, $\mu_{\rm p}$. In addition, we tested two estimates for the amplitude modification in the g-mode cavity: radiative damping (i.e., $\mu_{\rm g} = \mu_{\rm g,rad}$) and complete dissipation (i.e., $\mu_{\rm g} = \infty$). This results in a total of six combinations of $\mu_{\rm p}$ and $\mu_{\rm g}$ for each model.
We then integrated the power over the frequency ranges in Table \ref{tab: l segments} to obtain the contribution of the degree $\ell_{\rm a}$ to the integrated power associated with the degree $\ell_{\rm b}$. We denote this contribution of $\ell_{\rm a}$ to $\ell_{\rm b}$ as $C^{\ell_{\rm a}\rightarrow\ell_{\rm b}}$. 

A meaningful comparison between predicted visibilities and observations is only possible if the efficiency of the damping process in the g-mode cavity is comparable to the actual damping processes in the stars. Therefore, we divide the observed population of RGB stars into a “normal” sample containing stars with $(\mathcal{V}_{\ell=1}^2)^{\rm Stello} \geq 0.7$ and a “depressed” sample containing stars with $(\mathcal{V}_{\ell=1}^2)^{\rm Stello} < 0.7$. The choice of the visibility threshold does not significantly affect our results. The two samples are shown in Fig. \ref{fig: visibility stello}. For the analysis performed in this section, we restrict ourselves to the $\nu_{\rm max}$ range in which the visibilities of the stars in the “normal” and “depressed” samples appear as two distinct populations, which means that we only consider stars within $70\ \mu{\rm Hz} \lesssim \nu_{\rm max}\lesssim 240\ \mu{\rm Hz}$. Since no clear mass trend as a function of $\nu_{\rm max}$ was observed in the visibility \citep[e.g.,][]{stello+16a,stello+16b}, we include all stars irrespective of their mass to maintain a larger sample size.

We compare the stars in the “normal” sample with our predictions using radiative damping (i.e., $\mu_{\rm g} = \mu_{\rm g,rad}$) and the stars in the “depressed” sample using complete dissipation (i.e., $\mu_{\rm g} = \infty$).
For each model of a given evolutionary track, we calculate the average observed visibility $\langle\mathcal{V}_{\ell}^2\rangle^{\rm Stello}_{\nu_i}$ as the median visibility of the observed stars in the sample under consideration.
To do so, we only consider stars whose value of $\nu_{\rm max}$ is closest to the $\nu_{\rm max}$ value of the model (denoted as $\nu_i$) compared to the other models on the same track (see magenta lines in Fig. \ref{fig: visibility stello}). We then determine the optimal values of $\mathcal{S}_{\ell=1}$ and $\mathcal{S}_{\ell=2}$ by minimizing the parameter $D$, which we define as
\begin{gather}
    D = \sum_{\nu_i \in \{\nu_{\rm max} \}} \frac{n_{\nu_i}}{n_{\rm tot}} \left[ \left( \frac{C^{0\rightarrow1}_{\nu_i} + \mathcal{S}_{\ell=1} C^{1\rightarrow1}_{\nu_i} + \mathcal{S}_{\ell=2} C^{2\rightarrow1}_{\nu_i}}
    {C^{0\rightarrow0}_{\nu_i} + \mathcal{S}_{\ell=1} C^{1\rightarrow0}_{\nu_i} + \mathcal{S}_{\ell=2} C^{2\rightarrow0}_{\nu_i}} - \left\langle\mathcal{V}_{\ell=1}^2\right\rangle^{\rm Stello}_{\nu_i} \right)^2 \right. \notag\\ \qquad
    \left. + \left( \frac{C^{0\rightarrow2}_{\nu_i} + \mathcal{S}_{\ell=1} C^{1\rightarrow2}_{\nu_i} + \mathcal{S}_{\ell=2} C^{2\rightarrow2}_{\nu_i}}
    {C^{0\rightarrow0}_{\nu_i} + \mathcal{S}_{\ell=1} C^{1\rightarrow0}_{\nu_i} + \mathcal{S}_{\ell=2} C^{2\rightarrow0}_{\nu_i}}- \left\langle\mathcal{V}_{\ell=2}^2\right\rangle^{\rm Stello}_{\nu_i} \right)^2 \right].
\end{gather}
In this expression, $\{\nu_{\rm max} \}$ is the set of $\nu_{\rm max}$-values of the models of the selected evolutionary track, $C^{\ell_{\rm a}\rightarrow\ell_{\rm b}}_{\nu_i}$ is the contribution of the spherical degree $\ell_{\rm a}$ to $\ell_{\rm b}$ of the model with $\nu_{\rm max} = \nu_i$ (see above), $n_{\nu_i}$ is the number of stars used to calculate $\langle\mathcal{V}_{\ell}^2\rangle^{\rm Stello}_{\nu_i}$, and $n_{\rm tot}$ is the total number of stars observed by \citet{stello+16a,stello+16b} in the sample under consideration.

As a sanity check, we perform the same analysis again, this time calculating $\langle\mathcal{V}_{\ell}^2\rangle^{\rm Stello}_{\nu_i}$ based on observed stars with a similar value of $\Delta\nu$ as the model, rather than considering stars with a similar $\nu_{\rm max}$. We examine the $\Delta\nu$-range $7\ \mu{\rm Hz} \lesssim \Delta\nu\lesssim 20\ \mu{\rm Hz}$. Apart from this, the procedure remains unchanged.

\subsection{Estimates for $\mathcal{S}_{\ell=1}$ and $\mathcal{S}_{\ell=2}$}

\begin{table}[]
    \caption{Estimates of the spatial response $\mathcal{S}_{\ell}$.}
    \centering
    \begin{tabular}{cccccc}
    \hline\hline
    $M_\star / M_\odot$ & $\mu_{\rm g}$ & \multicolumn{2}{c}{$\mathcal{S}_{\ell=1}$} & \multicolumn{2}{c}{$\mathcal{S}_{\ell=2}$} \\
    \multicolumn{2}{c}{} & ($\nu_{\rm max}$) & ($\Delta\nu$) & ($\nu_{\rm max}$) & ($\Delta\nu$) \\
    \hline
    1 & $\mu_{\rm g,rad}$ & $1.47^{+0.03}_{-0.02}$ & $1.47^{+0.03}_{-0.03}$ & $0.69^{+0.01}_{-0.01}$ & $0.69^{+0.01}_{-0.01}$ \\
    1.25 & $\mu_{\rm g,rad}$ & $1.48^{+0.04}_{-0.03}$ & $1.48^{+0.04}_{-0.03}$ & $0.70^{+0.01}_{-0.01}$ & $0.70^{+0.01}_{-0.01}$ \\
    1.5 & $\mu_{\rm g,rad}$ & $1.47^{+0.05}_{-0.03}$ & $1.47^{+0.05}_{-0.03}$ & $0.70^{+0.02}_{-0.01}$& $0.70^{+0.02}_{-0.01}$ \\
    \hline
    1 & $\infty$ & $1.63^{+1.49}_{-0.51}$ & $1.26^{+1.16}_{-0.40}$ & $0.68^{+0.15}_{-0.06}$ & $0.62^{+0.12}_{-0.05}$ \\
    1.25 & $\infty$ & $1.62^{+1.53}_{-0.52}$ & $1.38^{+1.30}_{-0.44}$ & $0.69^{+0.17}_{-0.07}$ & $0.64^{+0.14}_{-0.06}$ \\
    1.5 & $\infty$ & $1.44^{+1.36}_{-0.46}$ & $1.34^{+1.27}_{-0.43}$ & $0.68^{+0.18}_{-0.07}$ & $0.66^{+0.17}_{-0.07}$ \\
    \hline
     & & \multicolumn{2}{c}{1.47} & \multicolumn{2}{c}{0.70} \\
    \hline
    \end{tabular}
    \tablefoot{The spatial responses $\mathcal{S}_{\ell=1}$ and $\mathcal{S}_{\ell=2}$ are given in two columns: the left column shows the values estimated from sampling the observed data over $\nu_{\rm max}$, and the right column shows the values estimated from sampling over $\Delta\nu$ (see main text). The given value of $\mathcal{S}_{\ell}$ corresponds to the intermediate value of the radial mode linewidth $\Gamma_0$, while the deviation of this value from the upper (lower) value of $\Gamma_0$ is indicated by the negative (positive) errors. The values of $\mathcal{S}_{\ell}$ in the bottom row are those used in this work. A visualization of the data presented in this table can be seen in Fig. \ref{fig: spatial response plot}.}
    \label{tab: spatial response}
\end{table}

\begin{figure}[]
    \centering
    \resizebox{\hsize}{!}{\includegraphics{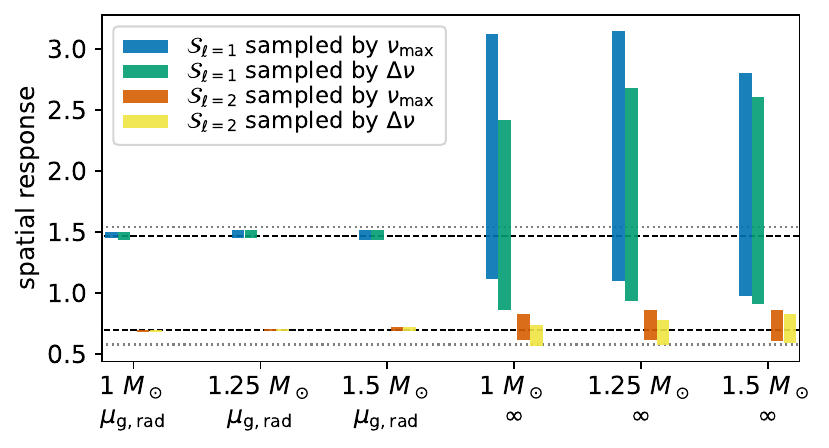}}
    \caption{Estimates for the spatial response for different stellar masses and efficiencies of the damping process in the g-mode cavity. The same data are also listed in Table \ref{tab: spatial response}. Vertical bars indicate the spread in the radial mode linewidth. Dashed black lines mark the vales selected for this study and dotted gray lines mark the theoretical values for $T_{\rm eff} = 4800$ K \citep{ballot+11}. Colors correspond to the spherical degree and whether the comparison was done for similar values of $\nu_{\rm max}$ or $\Delta\nu$.}
    \label{fig: spatial response plot}
\end{figure}

\begin{figure}[]
    \centering
    \resizebox{\hsize}{!}{\includegraphics{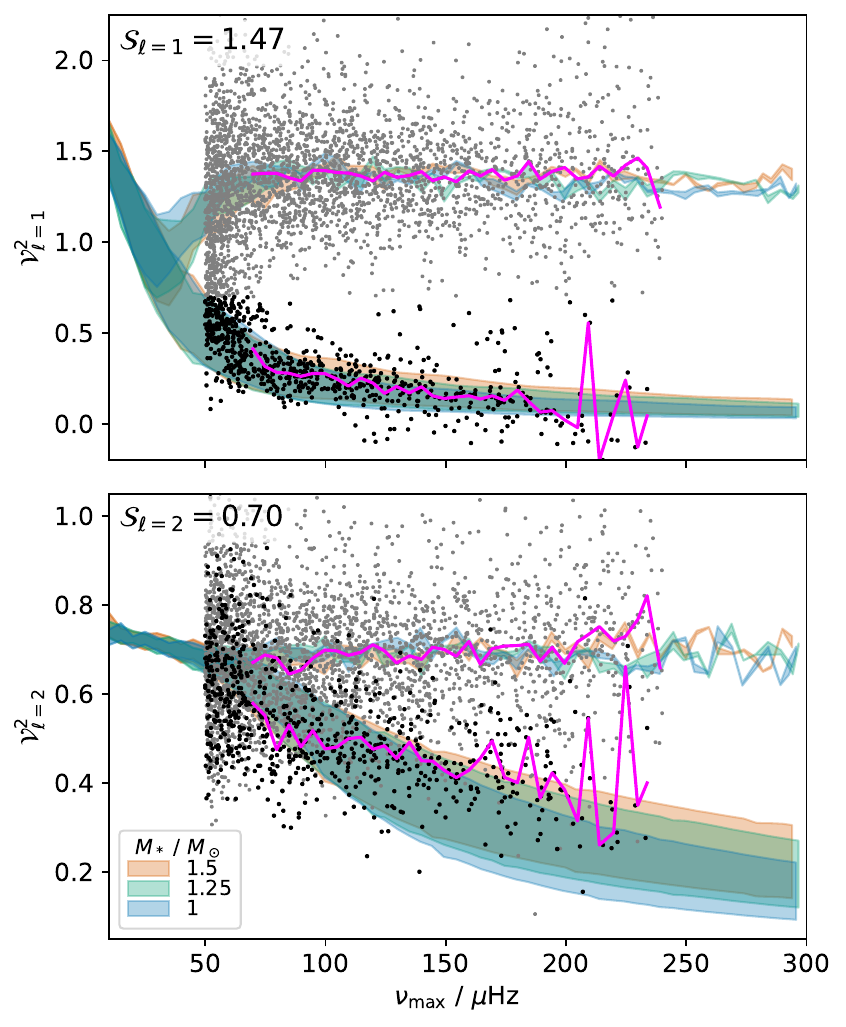}}
    \caption{Observable visibility of dipole modes {\it (upper panel)} and quadrupole modes {\it (lower panel)} as a function of the frequency of maximum oscillation power. Stars evolve from right to left. The shaded areas indicate the range of possible visibilities corresponding to different values of radial mode linewidth and are color-coded according to the stellar mass. In both panels, the upper shaded areas correspond to radiative damping (i.e., $\mu_{\rm g} = \mu_{\rm g,rad}$), while the lower areas correspond to complete suppression of oscillations in the g-mode cavity (i.e., $\mu_{\rm g} = \infty$). The points correspond to the visibilities observed by \citet{stello+16a,stello+16b}. Gray points indicate the normal stars and black points the depressed dipole mode stars. In magenta, we show the average (median) observed visibilities $\langle\mathcal{V}_{\ell}^2\rangle^{\rm Stello}$ sampled by $\nu_{\rm max}$ used to constrain the values of $\mathcal{S}_{\ell=1}$ and $\mathcal{S}_{\ell=2}$ for the track with $M_\star = 1.25\ M_\odot$.}
    \label{fig: visibility stello}
\end{figure}

We show our estimates of $\mathcal{S}_{\ell=1}$ and $\mathcal{S}_{\ell=2}$ in Table \ref{tab: spatial response}. In addition, we show a visualization of the same data in Fig. \ref{fig: spatial response plot}. The spread in the inferred values of $\mathcal{S}_{\ell}$ for different $\Gamma_0$ is much larger for the “depressed” sample than for the "normal" sample. 
This is likely because when $\mu_{\rm g} = \infty$, the visibility depends more strongly on $\Gamma_0$ than when $\mu_{\rm g}$ is low \citep[e.g.,][]{fuller+15, takata16b, mosser+17, muller+26}.
Furthermore, the damping rate of the stars in the “depressed” sample is high but probably finite \citep{mosser+17,muller+26}. Therefore, our assumption of complete dissipation of mode energy in the g-mode cavity is an approximation, and deviations from it could manifest themselves as a larger spread in $\mathcal{S}_{\ell}$ for different $\Gamma_0$. 

Taking into account the possible values of $\mathcal{S}_{\ell}$ from Table \ref{tab: spatial response}, we set $\mathcal{S}_{\ell=1} = 1.47$ and $\mathcal{S}_{\ell=2} = 0.70$ in this study. These values were determined as weighted averages and are consistent with the ranges of $\mathcal{S}_{\ell}$ for all considered stellar masses, both for the “normal” and the “depressed” sample. Our estimate for $\mathcal{S}_{\ell=1}$ is higher than the average dipole mode visibility of the normal stars estimated by \citet{stello+16a,stello+16b} (1.35), suggesting that neglecting gravity-dominated mixed modes that are far from the position of the nominal p-mode does indeed cause a significant bias in the calculation of the visibility. In particular, our estimate is closer to the theoretical value when assuming energy equipartition of ${\sim}1.54$ \citep{ballot+11} than the previous estimates.
However, our estimate of $\mathcal{S}_{\ell=2}$ is comparable to the average normal quadrupole mode visibility of \citet{stello+16a,stello+16b} (0.69) and therefore does not improve the agreement with the theoretical value assuming energy equipartition of ${\sim}0.58$. 

Deviations of the measured values of the bias-corrected $\mathcal{S}_{\ell}$ compared to the theoretical estimates of \citet{ballot+11} might suggest that the energy equipartition is not fully realized in real RGB stars. This may be caused by an $\ell$-dependence of the excitation and damping processes in the p-mode cavity.
For the dipole modes, the relative difference between the measured value of $\mathcal{S}_{\ell=1}$ and the theoretical value is roughly $-5$\%. For the quadrupole modes, the relative difference is significantly larger at almost $+20$\%.
If significant, these discrepancies could be due to the multipole modes being more or less energetic than the radial modes, but they could also be influenced by artificial effects that arose during the analysis of the stellar light curves. A prime candidate for this is the background correction. Although the background is fitted individually for each star, \citet{stello+16a,stello+16b} applied the same general method to all stars in their sample. A systematic trend toward lower or higher visibilities due to the residuals of the background correction is therefore possible and could contribute to the measurements of $\mathcal{S}_{\ell}$.
A dedicated observational study estimating the visibility using different background correction methods for the same sample of stars would be of interest to better assess the impact of the background correction on $\mathcal{S}_{\ell}$.

\subsection{Comparison to observations}

In Fig. \ref{fig: visibility stello}, we show the observable visibilities $\mathcal{V}^2_{\ell=1}$ and $\mathcal{V}^2_{\ell=2}$ predicted for radiative damping (i.e., $\mu_{\rm g} = \mu_{\rm g,rad}$) and complete loss of energy in the g-mode cavity (i.e., $\mu_{\rm g} = \infty$) and compare them to the visibilities measured for real stars, $(\mathcal{V}_\ell^2)^{\rm Stello}$\footnote{Note that our calculation of visibility is based on asymptotic theory, which assumes that the wavelength of the oscillations is small compared to the length scale of the variation in the stellar background structure. This is not the case for p-modes in evolved RGB stars (i.e., stars with low $\nu_{\rm max}$). Therefore, our predictions for visibility at the lower end of $\nu_{\rm max}$ in Fig. \ref{fig: visibility stello} should be treated with appropriate caution.}.
Figure \ref{fig: visibility stello} shows that the median value of the observed visibilities of the dipole modes, $\langle\mathcal{V}_{\ell=1}^2\rangle^{\rm Stello}$, is well reproduced by our predictions. The scattering of the observed visibilities around the median value is mainly caused by the stochastic nature of the modes.
For the quadrupole modes, the observed visibilities, $(\mathcal{V}_{\ell=2}^2)^{\rm Stello}$, also largely follow the predicted behavior. However, we expect a steeper trend in the evolution of visibility for the “depressed” sample compared to what is observed. A similar conclusion was drawn by \citet{stello+16b}. 
Assuming complete dissipation of the mode energy in the g-mode cavity, the (theoretical) visibility reduces to an expression that depends on $\Delta\nu$, $\Gamma_0$, and $q_\ell$ \citep{mosser+17}:
\begin{gather}
    V^2_\ell = \left(1 - \ln\left(1 - \frac{4q_\ell}{(1+q_\ell)^2}\right) \frac{\Delta\nu}{2 \pi \Gamma_0} \right)^{-1}.
    \label{eq: visibility full suppression}
\end{gather}
Since $\Delta\nu$ is presumably well constrained by the observations and the behavior of the dipole modes is adequately reproduced using $\Gamma_0$, the surprising behavior of the “depressed” branch in the visibility of the quadrupole modes could be attributed to an unexpected dependence of $q_{\ell=2}$ on $\nu_{\rm max}$. Alternatively, if the damping caused by interaction with convection turns out to be $\ell$-dependent, it might also lead to such behavior.

\section{Theoretical versus observable visibilities} \label{SECT: Theoretical versus observable visibilities}

\begin{figure*}[]
    \centering
    \resizebox{.49\hsize}{!}{\includegraphics{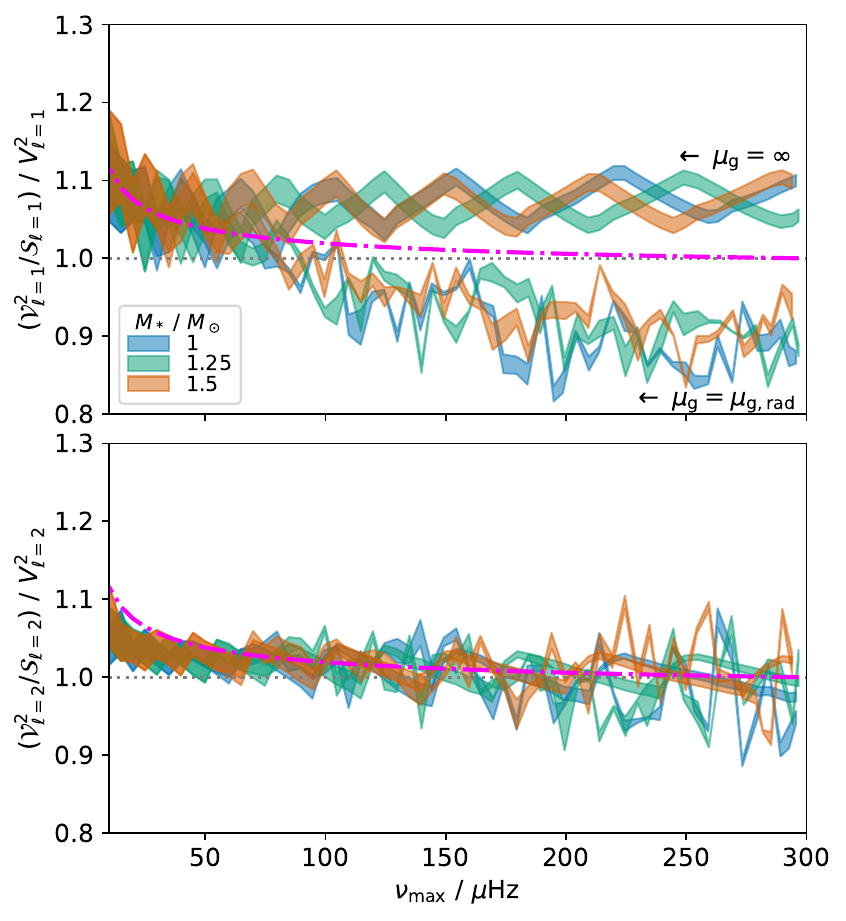}}
    \resizebox{.49\hsize}{!}{\includegraphics{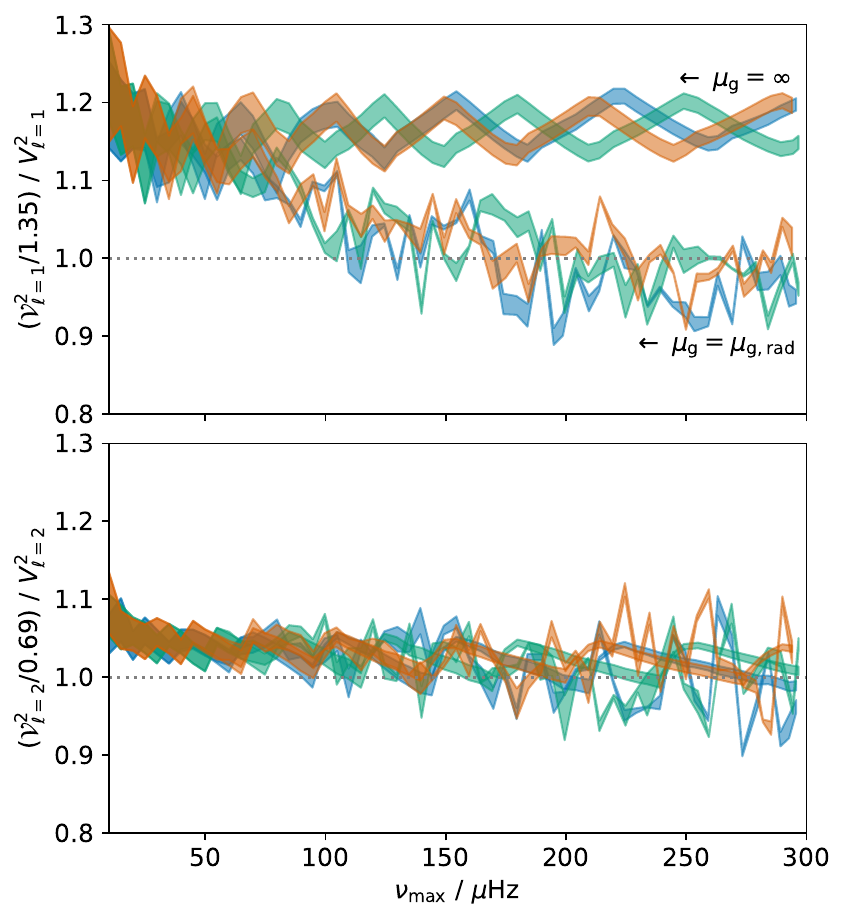}}
    \caption{Ratio between observable and theoretical visibility as a function of the frequency of maximum oscillation power for the dipole modes {\it (top row)} and the quadrupole modes {\it (bottom row)}. 
    The shaded areas correspond to the values allowed by the range of radial mode linewidths, and the colors indicate different stellar masses.
    In the \textit{left column}, the observable visibilities are normalized by $\mathcal{S}_{\ell}$ to allow comparison with the theoretical visibility. In the \textit{right column}, the observable visibilities are instead normalized by the average visibility of the normal stars determined by \citet{stello+16a,stello+16b}.
    For the damping process in the core, either radiative damping or complete dissipation of the energy entering the g-mode cavity was assumed. The type of damping used is indicated by the text in the top row. It is not indicated in the bottom row, as the two damping prescriptions yield comparable results for the quadrupole modes.
    The magenta dashed-dotted lines shown in the left column are the inverse of Eq. \ref{eq: contribution radial to radial} normalized for the model with $\nu_{\rm max} \approx 300\ \mu$Hz using the intermediate value of $\Gamma_0$.}
    \label{fig: comparison theo obs visibility}
\end{figure*}

\begin{figure*}[]
    \centering
    \resizebox{\hsize}{!}{\includegraphics{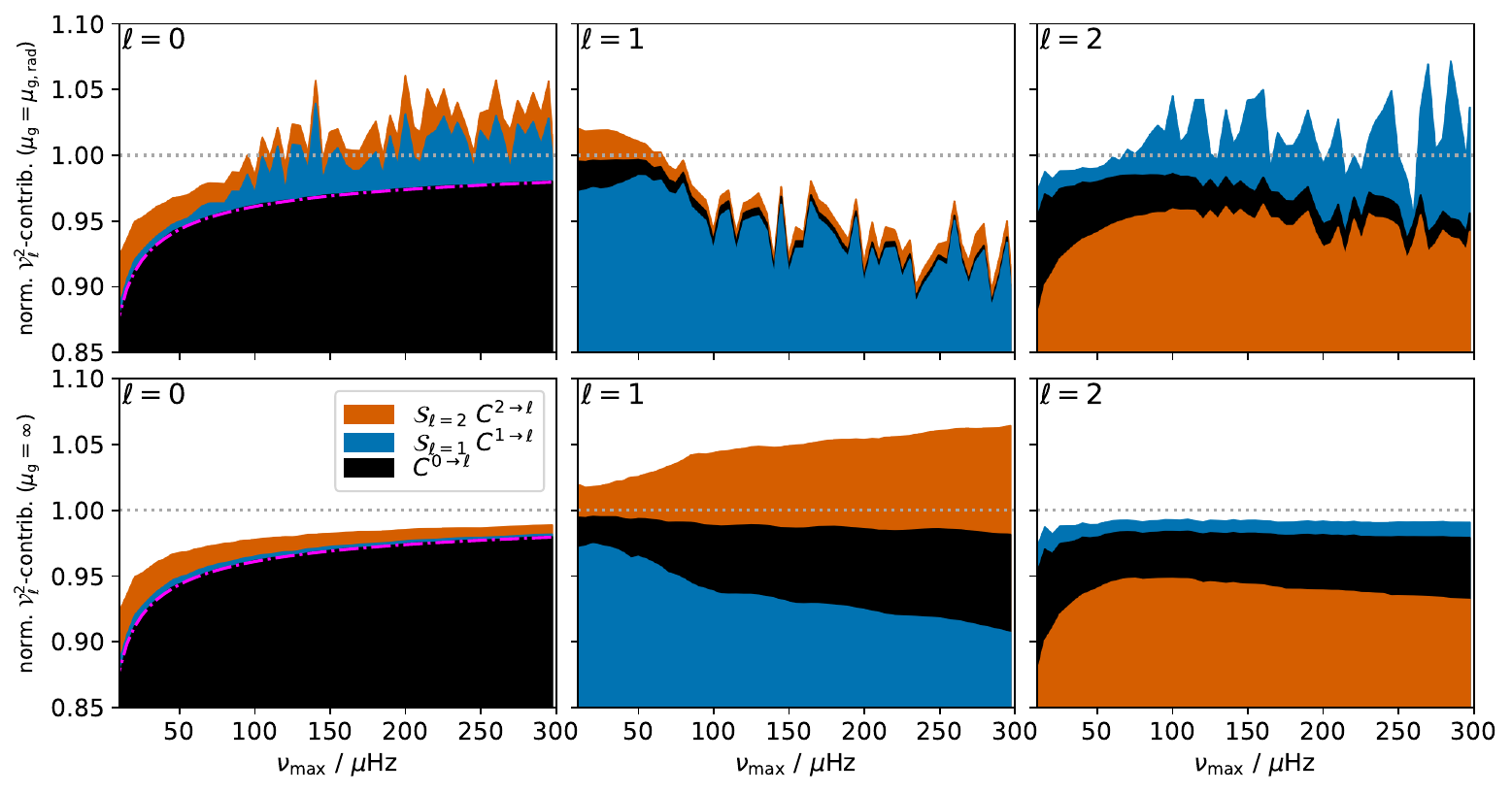}}
    \caption{Normalized sum of the contributions to the integrated power spectrum of the degree $\ell$ (i.e., $(C^{0\rightarrow\ell} + \mathcal{S}_{\ell=1} C^{1\rightarrow\ell} + \mathcal{S}_{\ell=2} C^{2\rightarrow\ell})/\mathcal{N}_\ell$) as a function of the frequency of the maximum oscillation power for the evolutionary track with $M_\star = 1.25\ M_\odot$ and the intermediate value of the radial mode linewidth. In the top row, we used radiative damping, and in the bottom row, all the energy entering the g-mode cavity is dissipated. In the columns, we show the different degrees that the contributions correspond to from an observational perspective. The colors indicate the degree to which the contributions theoretically correspond. The magenta dashed-dotted line has been computed with Eq. \ref{eq: contribution radial to radial}. The normalization factor is the total integrated power spectrum of the observed $\ell$, given by $\mathcal{N}_\ell = \mathcal{S}_{\ell} \, (C^{\ell\rightarrow0} + C^{\ell\rightarrow1} + C^{\ell\rightarrow2} + C^{\ell\rightarrow3})$.}
    \label{fig: visibility contributions}
\end{figure*}

In this section, we compare the theoretical visibility, $V_\ell^2$, with the observable visibility, $\mathcal{V}_\ell^2$. The observable visibility differs from the theoretical one due to the contribution of the spatial response $\mathcal{S}_{\ell}$, the multiplication by the Gaussian envelope, and the integration of the power spectrum over $\ell$-dependent segments instead of the entire frequency range (see Sect. \ref{sect: visibility}). The latter leads to the aforementioned bias, which can cause the gravity-dominated modes to be neglected when calculating the visibility of their corresponding $\ell$ and instead contribute to the visibility of a different degree. In Fig. \ref{fig: comparison theo obs visibility}, we show the ratio between the observable and theoretical visibility, where the observable visibility is normalized by $\mathcal{S}_{\ell}$.

Focusing first on the quadrupole modes, we see that $\mathcal{V}_{\ell=2}^2$ scatters around the theoretical value at higher values of $\nu_{\rm max}$. This can be understood by looking at Fig. \ref{fig: visibility contributions}, where we show the contribution of modes of different $\ell$ to the visibility of their respective and other degrees. In this figure, the upper left and right panels show that the scatter in the visibility ratio of the quadrupole modes is mostly caused by the occasional contribution of gravity-dominated dipole modes to the frequency segments corresponding to the radial and quadrupole modes. Since the number of dipole modes present in the frequency segment of the quadrupole modes varies with $\Delta\nu$ and $\Delta\Pi_{\ell=1}$, this contribution differs for each stellar model, resulting in the spiky appearance of the contributions of the gravity-dominated dipole modes to the visibilities. This spikiness fades out with decreasing coupling strength, as it causes the gravity-dominated modes to carry less power. The contribution of the quadrupole mode to the visibilities also show spiky features for the same reason, but they are less prominent due to their smaller coupling strength (i.e., $q_{\ell=2} < q_{\ell=1}$).

\subsection{Bias on the late RGB}

\begin{figure}[]
    \centering
    \resizebox{\hsize}{!}{\includegraphics{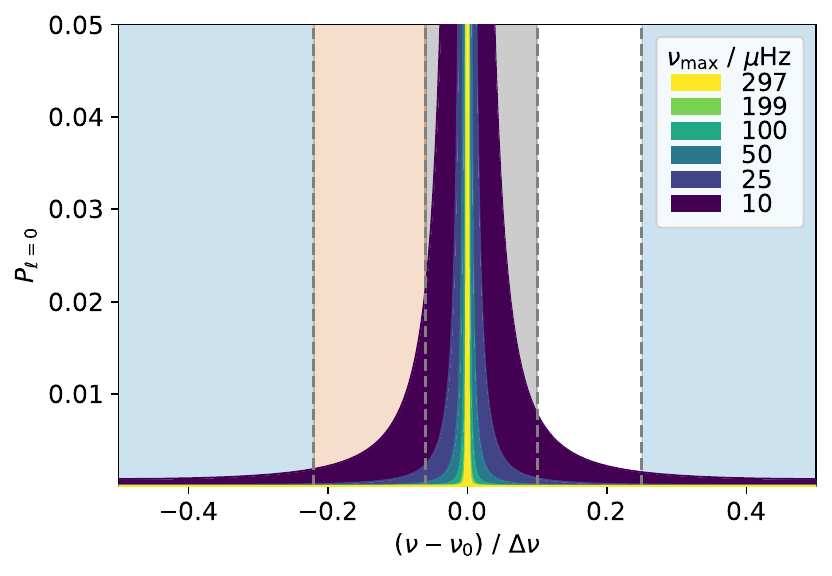}}
    \caption{Normalized power spectrum of a radial mode centered at $\nu_0$ as a function of frequency, color-coded according to different points in evolution along the RGB for the evolutionary track with $M_\star = 1.25\ M_\odot$ and the intermediate value of the radial mode linewidth. The shaded areas and vertical dashed lines show the frequency segments used for calculating the observable visibility (black: $\ell=0$, blue: $\ell=1$, red: $\ell=2$; see Table \ref{tab: l segments}).}
    \label{fig: radial mode}
\end{figure}

Figure \ref{fig: comparison theo obs visibility} shows that $\mathcal{V}_{\ell=2}^2$ overestimates the visibility by up to $10\%$ at lower $\nu_{\rm max}$ values. The left-hand panels of Fig. \ref{fig: visibility contributions} show that this increase in visibility is caused by the lower value of the integrated power in the frequency segments of the radial modes. This is because the radial modes are no longer captured by their corresponding frequency segments, as these segments become narrower with decreasing $\Delta\nu$ (see Table \ref{tab: l segments}). For this reason, the wings of the radial mode peaks are not taken into account when integrating the power spectrum over the corresponding frequency segments, which increases $\mathcal{V}_\ell^2$ independently of $\ell$. This is shown in Fig. \ref{fig: radial mode}.
A similar increase in the visibility ratio at lower $\nu_{\rm max}$ can also be observed for the dipole modes in Fig. \ref{fig: comparison theo obs visibility}. This is consistent with the observations of \citet{dreau+21}, who report that the visibility of very evolved RGB stars increases with decreasing $\nu_{\rm max}$.

Since the radial modes are Lorentzian peaks and not mixed, their contribution to $\mathcal{V}_{\ell=0}^2$ can be approximated analytically using the anti-derivative of a Lorentz function with hight unity and full width at half maximum $\Gamma_0$ centered around $\nu_0$:
\begin{gather}
    C^{0\rightarrow0} \approx \frac{1}{\pi}\left[\arctan\left( \frac{2 \left(\nu_{\ell=0}^{\rm high}-\nu_0 \right)}{\Gamma_0} \right) - \arctan\left( \frac{2\left(\nu_{\ell=0}^{\rm low}-\nu_0 \right)}{\Gamma_0} \right) \right] \notag\\
    \qquad \approx 1 -\left(\frac{1}{0.1} + \frac{1}{0.06}\right) \frac{\Gamma_0}{2\pi\Delta\nu} \approx  1 -4.24\, \frac{\Gamma_0}{\Delta\nu}.
    \label{eq: contribution radial to radial}
\end{gather}
In the second approximation, we used that $\Gamma_0 / \Delta\nu$ is small. Equation \ref{eq: contribution radial to radial} is shown as the magenta dashed-dotted line in the left panels of Fig. \ref{fig: visibility contributions} and agrees excellently with the numerical results. Both $\Delta\nu$ and $\Gamma_0$ are accessible in most asteroseismic observations, which means that Eq. \ref{eq: contribution radial to radial} can be used as a scaling relation to get a rough estimate for the severity of the bias on the late RGB (see magenta dashed-dotted lines in Fig. \ref{fig: comparison theo obs visibility}). In addition, this bias should also be significant for red clump stars, as these stars have small $\Delta\nu$ and a relatively wide range of $\Gamma_0$ \citep[e.g.,][]{vrard+18}.

\subsection{Biases on the early RGB}

Considering the visibility ratio of the dipole modes in Fig. \ref{fig: comparison theo obs visibility}, we see that at higher $\nu_{\rm max}$, the observable visibility, $\mathcal{V}_{\ell=1}^2$, underestimates the theoretical value when using radiative damping, but overestimates it when the all mode energy leaking into the g-mode cavity is lost. These trends can be explained using Fig. \ref{fig: visibility contributions}.

\subsubsection{Normal stars}

In the case of normal stars with radiative damping (top row of Fig. \ref{fig: visibility contributions}), $\mathcal{V}_{\ell=1}^2$ is lower because the gravity-dominated dipole modes that contribute to the integrated power spectrum of the radial and quadrupole modes are absent in the frequency segments of the dipole mode. This effect weakens as the star evolves due to the decreasing $q_{\ell=1}$, since the coupling strength determines how much energy is contained in the gravity-dominated mixed modes compared to the pressure-dominated modes. This is also why this effect is less pronounced for the quadrupole modes, as their coupling strength is significantly smaller than that of the dipole modes.

\subsubsection{Depressed dipole mode stars}

In the case of depressed dipole mode stars with complete dissipation of the mode energy in the g-mode cavity, the dipole mode visibility is so low that the contributions caused by the wings of the radial and quadrupole mode peaks significantly increase $\mathcal{V}_{\ell=1}^2$ compared to the theoretical value. This effect weakens during the course of stellar evolution due to the increasing intrinsic visibility of the dipole modes (which is caused by the decreasing value of $q_{\ell=1}$).
Note that the oscillating behavior of the visibility ratio, which can be clearly seen in the upper panels of Fig. \ref{fig: comparison theo obs visibility}, is caused by multiplication with the Gaussian envelope during the calculation of $\mathcal{V}_{\ell=1}^2$.
This is due to the variation in the position of the peaks relative to the center of the Gaussian envelope.

\subsection{Normalization by the average visibility of the normal stars}

In Fig. \ref{fig: comparison theo obs visibility}, we show in the right column the visibility ratio where $\mathcal{V}_{\ell=1}^2$ is normalized to 1.35 instead of our estimate of $\mathcal{S}_{\ell=1}$ in the upper panel and the visibility ratio where $\mathcal{V}_{\ell=2}^2$ is normalized to 0.69 instead of our estimate of $\mathcal{S}_{\ell=2}$ in the lower panel. These values were determined by \citet{stello+16a,stello+16b} as the average observed visibilities of the normal stars. In this case, the theoretical and observable dipole mode visibilities are comparable for higher and intermediate values of $\nu_{\rm max}$ when using radiative damping. This is to be expected, since 1.35 is the average observed visibility, as mentioned earlier. However, doing so overestimates the dipole mode visibility by about $20\%$ in the case of radiative damping for lower $\nu_{\rm max}$ and for the entire evolution on the RGB when considering complete dissipation of the mode energy in the g-mode cavity. This analysis demonstrates that the measured visibilities of the depressed dipole modes stars should not be compared quantitatively with theoretical predictions without taking the biases discussed in this section into account.
On the other hand, the quadrupole modes are not strongly affected by the different normalization, since the average quadrupole mode visibility determined by \citet{stello+16a,stello+16b} is very close to our estimate of $\mathcal{S}_{\ell=2}$.

\section{Observable signatures of strong magnetic damping in the core} \label{SECT: Observable signatures of magnetic damping}

\begin{figure*}[]
    \centering
    \resizebox{\hsize}{!}{\includegraphics{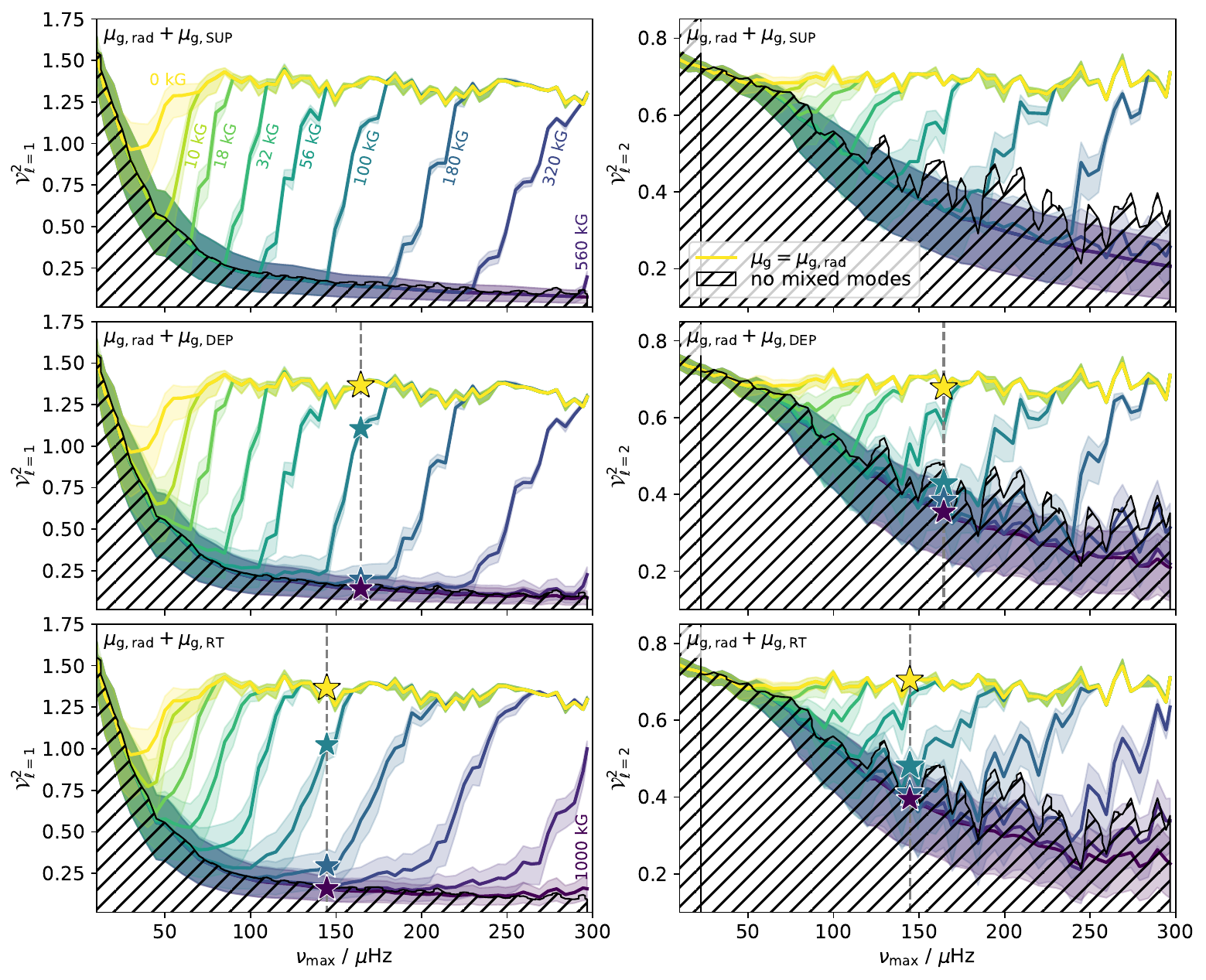}}
    \caption{Observable dipole mode visibility {\it (left column)} and quadrupole mode visibility {\it (right column)} as a function of frequency of maximum oscillation power for the evolutionary track with $M_\star = 1.25\ M_\odot$. {\it Rows} correspond to different assumptions for the energy loss caused by a strong magnetic field in the g-mode cavity (see Sect. \ref{sect: damping processes}). Colors indicate different values of the initial central field strength on the main sequence $B_{\rm MS}$ (see upper and bottom left panel). Damping due to the interaction with convection and radiative damping are always taken into account. Solid lines were calculated using the intermediate value of the radial mode linewidth $\Gamma_0$, and the shaded areas indicate the range of possible values permitted by the range of $\Gamma_0$ under consideration. Black hatched areas indicate where mixed modes cannot be detected. The hatched areas were calculated using the intermediate value of $\Gamma_0$ and should be compared to the solid lines (and not to the shaded areas). The vertical hatched band at low values of $\nu_{\rm max}$ in the right column indicates that the mixed signature of the quadrupole modes is never detectable. The PSDs of the models highlighted with a star symbol are shown in Figs. \ref{fig: PSD DEP} and \ref{fig: PSD RT}.}
    \label{fig: magnetic damping evolution}
\end{figure*}
\begin{figure}[]
    \centering
    \resizebox{\hsize}{!}{\includegraphics{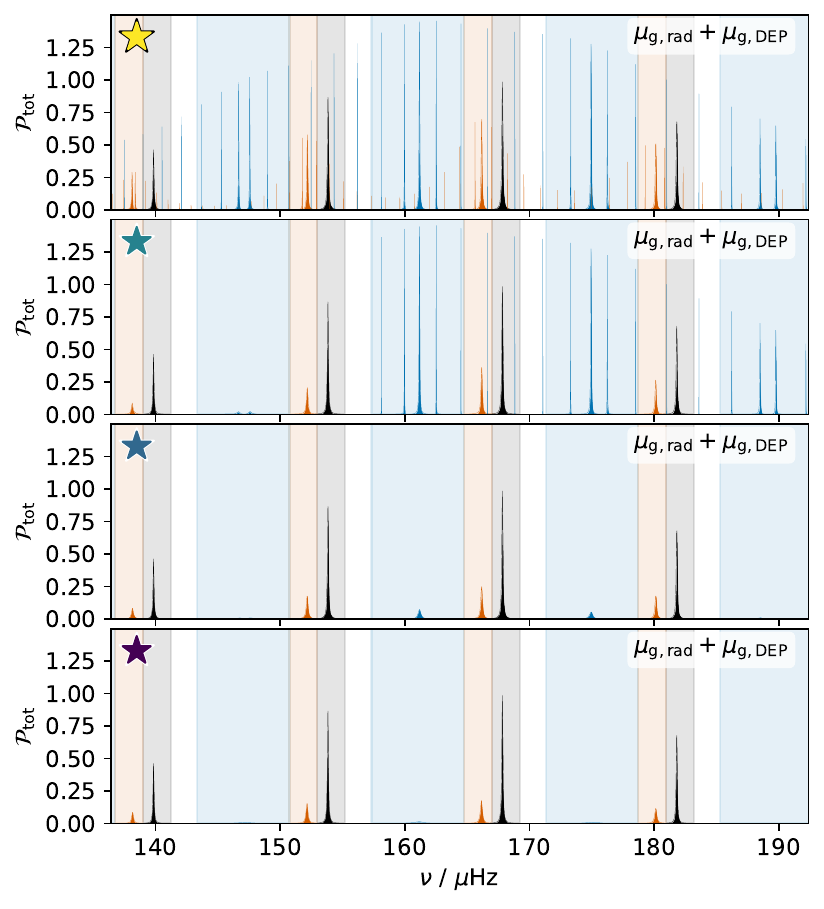}}
    \caption{Total power spectrum as a function of frequency, taking into account damping due to interaction with convection, radiative damping, and ad hoc depression of mixed modes (DEP) for the models marked with star symbols in Fig. \ref{fig: magnetic damping evolution}. The black peaks correspond to the radial modes, the blue peak to the dipole modes, and the red peaks to the quadrupole modes. The shaded areas show the frequency segments used for calculating the observable visibility (see Table \ref{tab: l segments}).}
    \label{fig: PSD DEP}
\end{figure}
\begin{figure}[]
    \centering
    \resizebox{\hsize}{!}{\includegraphics{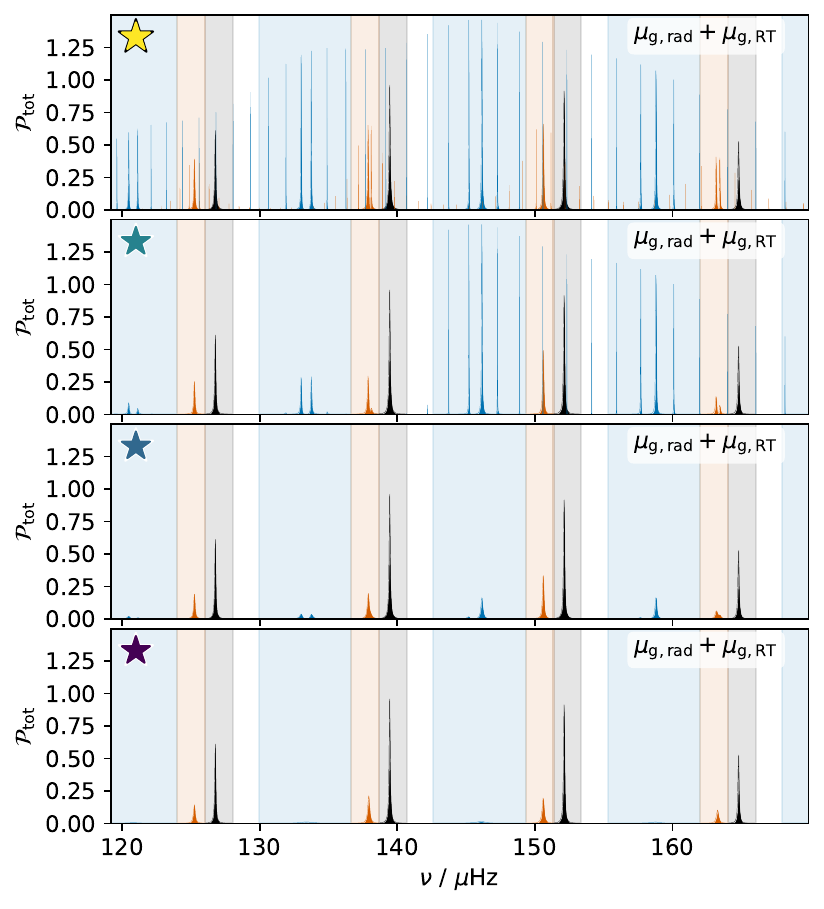}}
    \caption{Same as Fig. \ref{fig: PSD DEP}, now using the predictions from the ray tracing analysis of magneto-gravity waves (RT) and for a different stellar model.}
    \label{fig: PSD RT}
\end{figure}

Here, we explore the observable signatures produced by the three different prescriptions for energy loss that could be caused by an internal magnetic field introduced in Sect. \ref{sect: damping processes} (i.e., SUP, DEP, and RT). We test a range of initial magnetic field strengths on the main sequence, $B_{\rm MS}$, (see Sect. \ref{sect: initialization}) and evolve the field strength using the conservation of magnetic flux. We always take into account the damping due to interaction with convection $\mu_{\rm p}$ and radiative damping $\mu_{\rm g,rad}$. Since the results for all our evolutionary tracks are qualitatively comparable, we discuss our results based on the track with $M_\star = 1.25\ M_\odot$.
In Fig. \ref{fig: magnetic damping evolution}, we show that, taking into account the energy loss caused by interaction with the magnetic field, visibility mostly falls on (or close to) one of two branches, which correspond to the normal RGB stars and the depressed dipole modes stars. These branches are connected by a short transition event that occurs when the magnetic field strength approaches the critical field strength. Thus, the transition occurs at higher values of $\nu_{\rm max}$ for higher values of $B_{\rm MS}$. As expected, the transition sets in earlier for the quadrupole modes than for the dipole modes (see Eq. \eqref{eq: B crit}). At small values of $\nu_{\rm max}$, the two branches merge because the coupling strength decreases as the star ascends the RGB.

Apart from visibility, the presence or absence of dipolar mixed modes is another observational feature that provides information about the damping rate of the damping process in the g-mode cavity. In the observed population of depressed dipole modes stars, both stars without \citep{stello+16a,stello+16b} and stars with signs of mixed dipolar modes \citep{mosser+17} were identified. In this study, we used the detection criterion defined by \citet{muller+26} to investigate whether the mixed signature of the dipole modes is theoretically detectable in our models. It should be noted that due to the theoretical nature of this approach, which neglects observational effects such as residuals from background correction, stochastic excitation, and noise, the threshold shown in Fig. \ref{fig: magnetic damping evolution}, at which the mixed modes are no longer detectable, should be considered a lower limit. This means that although we expect to detect mixed modes according to the detection criterion, they may not be identifiable in the observations. Conversely, however, we can confidently predict that we will not expect to see any signs of mixed modes in observations if there are no detectable peaks according to our criterion. Furthermore, for the sake of simplicity, we neglect the influence of mixed mode multiplets and the inclination angle of the star onto the detectability of the mixed mode signature \citep[see][]{muller+26}.

The top row of Fig. \ref{fig: magnetic damping evolution} clearly shows that when the mode energy in the g-mode cavity is completely dissipated (SUP), the mixed signature of the dipole modes disappears as soon as the star reaches the branch that corresponds to the depressed dipole mode stars. Crucially, this is inconsistent with observations of depressed dipole mode stars that have very low dipole mode amplitudes but still show a clear mixed mode signature across several pressure radial orders \citep{mosser+17}. 
If, on the other hand, a small fraction of the mode energy can escape from the g- to the p-mode cavity, depressed dipole modes with a mixed character can appear in the PSD. This is shown in the middle and bottom rows of Fig. \ref{fig: magnetic damping evolution}, where we have used the energy loss prescriptions according to the ad hoc depression of mode energy (DEP) and the ray tracing analysis of magneto-gravity waves (RT).

This is further illustrated in Figs. \ref{fig: PSD DEP} and \ref{fig: PSD RT}, where we show synthetic PSDs using either the DEP or RT prescription for different initial magnetic field strengths. The second panels of these two figures show that modes with lower $\nu$ are more susceptible to damping, which is a direct consequence of the dependence of the critical field strength on $\nu$ (see Eq. \eqref{eq: B crit}). Interestingly, as highlighted in several other studies \citep[e.g.,][] {mosser+17,loi20a,deheuvels+23}, the observed PSD of the star KIC 6975038 \citep[see e.g., Figure 7 of] []{muller+25} shows precisely this behavior of normal dipole modes at higher $\nu$ and damped dipole modes at lower $\nu$. Stars with PSDs featuring trends similar to this could, in principle, provide an explanation for the RGB stars with intermediate dipole mode visibility that do not fall on either the “normal” or the “depressed” branch (see Fig. \ref{fig: visibility stello}).
Furthermore, Fig. \ref{fig: magnetic damping evolution} shows that even if the mixed signature of depressed dipole modes is initially observable when using the DEP or RT prescriptions, it eventually disappears as the star evolves on the RGB. In particular, this means that the DEP or RT prescriptions allow for both the presence or absence of damped dipolar mixed modes, therefore agreeing phenomenologically with the observations, contrary to the SUP prescription.

\section{Conclusions} \label{SECT: Conclusions}

In this work, we generated synthetic PSDs for numerical stellar models along the RGB to determine the dipole and quadrupole visibility both purely theoretically and based on the observational techniques that are used for actual PSDs. In particular, we have taken into account the damping of the oscillations due to the interaction with convection measured by the radial mode linewidth and its dependence on the evolution of the star \citep{vrard+18}, radiative damping in the core, as well as state-of-the-art asymptotic expressions for calculating the coupling strength of the dipole modes \citep {shibahashi79,takata16a}. In addition, we tested three different prescriptions for the energy loss that could be caused by the interaction of the oscillations with a strong internal magnetic field \citep{fuller+15,muller+25}.

First, we compared our predictions for the observable visibility with actual observations from \citet{stello+16a,stello+16b} and estimated the spatial response, $\mathcal{S}_{\ell}$, which takes into account $\ell$-dependent observational effects such as limb darkening. \citet{stello+16a,stello+16b} estimated the average visibility of the normal stars, which corresponds to this factor, to be ${\sim}1.35$ for the dipole modes and ${\sim}0.69$ for the quadrupole modes.
The novelty of our approach is that, based on our synthetic PSDs, we can take into account the observational biases that arise from integrating the power spectrum over $\ell$-dependent frequency segments instead of over the entire frequency range, and from the fact that the observable visibility is contaminated by the contribution of modes with different $\ell$.
According to our analysis, the spatial responses of the dipole and quadrupole modes are $\mathcal{S}_{\ell=1}=1.47$ and $\mathcal{S}_{\ell=2}=0.70$, respectively. While the value for the quadrupole modes remains comparable to previous estimates, we find that $\mathcal{S}_{\ell=1}$ was significantly underestimated due to the neglect of gravity-dominated mixed modes in the calculation of the observable visibility (see Sect. \ref{sect: Introduction}). Our estimate of $\mathcal{S}_{\ell=1}$ is closer to the theoretical value expected when assuming energy equipartition, which is ${\sim}1.54$ for dipole modes and ${\sim}0.58$ for quadrupole modes \citep{ballot+11}. It is possible that the remaining difference between our estimate and the expected value of $\mathcal{S}_{\ell=1}$ is caused by systematic errors introduced by, for example, the background correction. A dedicated observational study considering various methods for correcting the background would be required to definitively determine the extent to which energy equipartition between the oscillation modes is realized in RGB stars.

Next, we compared theoretical and observable visibilities to investigate the extent to which the aforementioned observational biases affect the visibility values reported by observational studies. 
When the observable dipole mode visibility is normalized to 1.35 \citep[i.e., the average dipole mode visibility of the normal stars;][]{stello+16a,stello+16b}, it is comparable to the theoretical visibility on the early RGB if no strong damping process takes place in the core. However, the observable visibilities are overestimated by up to ${\sim}$20\% on the late RGB. For the depressed dipole modes stars, the observable visibilities are overestimated by ${\sim}$20\% throughout the entire evolution of the RGB. Crucially, this means that theoretical estimates of the visibility assuming complete dissipation of mode energy in the g-mode cavity \citep[e.g.,][]{fuller+15,mosser+17} cannot be directly compared quantitatively with the observed visibilities.
If the observable dipole mode visibility is instead normalized by our estimate of $\mathcal{S}_{\ell=1}$ instead of 1.35, the normalized observable visibility decreases by ${\sim}$10\%, so that it is now underestimated for stars with normal multipole mode amplitudes on the early RGB.
The quadrupole mode visibility is less biased and is only overestimated on the late RGB.
We expect the observational biases to be even greater for red clump stars, as these stars tend to have a smaller $\Delta\nu$ and a larger $\Gamma_0$ than RGB stars. This will be investigated in the next article in this series of publications.

Finally, we tested three prescriptions that describe the energy loss caused by a strong internal magnetic field in the core of the star.
We argue that the magnetic field strength should approach the critical field strength at the center of the star (and not at the hydrogen-burning shell).
In particular, we showed that, regardless of the implementation used, the multipole mode visibility rapidly transitions from the branch of RGB stars with normal multipole mode amplitudes to the branch corresponding to the depressed dipole mode stars as the field strength approaches the critical field strength. Furthermore, we find that if the damping mechanism allows some of the mode energy to escape from the g-mode cavity, the mixed mode signature can be both present or absent when the stars are on the branch of the depressed dipole mode stars, which is necessary to agree with observations \citep{mosser+17}. This applies both to the damping prescription in which a constant fraction of the mode energy returns from the g-mode cavity (DEP) and to the prescription in which this fraction decreases continuously with increasing magnetic field strength (RT). Therefore, both prescriptions are phenomenologically consistent with the observations.
However, partial dissipation of mode energy caused by an internal magnetic field does seem to be inconsistent with the results of most theoretical studies \citep{fuller+15,lecoanet+17,rui+fuller23,david+26}.
Based on the results of ray tracing studies \citep{loi20a, muller+25}, we hypothesize that, by taking into account the inner turning point of the g-mode cavity, which is not considered in the studies predicting complete dissipation, a portion of the mode energy could be preserved and reflected during the interaction of the waves with a strong magnetic field (see Sect. \ref{sect: location of magnetic damping}). This calls for a global mode solution.

\begin{acknowledgements}
We thank the anonymous referee for their comments, which greatly improved the clarity of the article. We would also like to thank the authors of \citet{stello+16a,stello+16b} for making their data publicly available.
We acknowledge funding from the ERC Consolidator Grant DipolarSound (grant agreement \# 101000296). In addition, we acknowledge support from the Klaus Tschira Foundation.
\end{acknowledgements}

\bibliographystyle{aa}
\bibliography{ref}

@ARTICLE{deheuvels+26,
       author = {{Deheuvels}, S. and {Ballot}, J. and {Ligni{\`e}res}, F. and {Li}, G. and {Villate}, M.},
        title = "{Near-critical magnetic fields in Kepler red giants}",
      journal = {\aap},
         year = 2026,
        month = may,
       volume = {709},
          eid = {A224},
        pages = {A224},
          doi = {10.1051/0004-6361/202659344},
archivePrefix = {arXiv},
       eprint = {2604.09901},
 primaryClass = {astro-ph.SR},
       adsurl = {https://ui.adsabs.harvard.edu/abs/2026A&A...709A.224D}
}

@ARTICLE{crawford+25,
       author = {{Crawford}, Courtney L. and {Li}, Yaguang and {Huber}, Daniel and {Yu}, Jie and {Bedding}, Timothy R. and {Martell}, Sarah L. and {Montet}, Benjamin T. and {Stello}, Dennis and {Isaacson}, Howard and {Howard}, Andrew W. and {Fulton}, Benjamin J. and {Zhang}, Jingwen and {Polanski}, Alex S. and {Weiss}, Lauren M.},
        title = "{The highest mass Kepler red giants ─ II. Spectroscopic parameters, the amplitude─activity relation, and unexpected halo orbits}",
      journal = {\mnras},
         year = 2025,
        month = oct,
       volume = {542},
       number = {4},
        pages = {3289-3301},
          doi = {10.1093/mnras/staf1421},
archivePrefix = {arXiv},
       eprint = {2508.12585},
 primaryClass = {astro-ph.SR},
       adsurl = {https://ui.adsabs.harvard.edu/abs/2025MNRAS.542.3289C}
}

@ARTICLE{gehan+24,
       author = {{Gehan}, C. and {Godoy-Rivera}, D. and {Gaulme}, P.},
        title = "{Magnetic activity of red giants: Correlation between the amplitude of solar-like oscillations and chromospheric indicators}",
      journal = {\aap},
         year = 2024,
        month = jun,
       volume = {686},
          eid = {A93},
        pages = {A93},
          doi = {10.1051/0004-6361/202349008},
archivePrefix = {arXiv},
       eprint = {2401.13549},
 primaryClass = {astro-ph.SR},
       adsurl = {https://ui.adsabs.harvard.edu/abs/2024A&A...686A..93G}
}

@ARTICLE{gehan+22,
       author = {{Gehan}, Charlotte and {Gaulme}, Patrick and {Yu}, Jie},
        title = "{Surface magnetism of rapidly rotating red giants: Single versus close binary stars}",
      journal = {\aap},
         year = 2022,
        month = dec,
       volume = {668},
          eid = {A116},
        pages = {A116},
          doi = {10.1051/0004-6361/202245083},
archivePrefix = {arXiv},
       eprint = {2211.01026},
 primaryClass = {astro-ph.SR},
       adsurl = {https://ui.adsabs.harvard.edu/abs/2022A&A...668A.116G}
}

@ARTICLE{gaulme+20,
       author = {{Gaulme}, Patrick and {Jackiewicz}, Jason and {Spada}, Federico and {Chojnowski}, Drew and {Mosser}, Beno{\^\i}t and {McKeever}, Jean and {Hedlund}, Anne and {Vrard}, Mathieu and {Benbakoura}, Mansour and {Damiani}, Cilia},
        title = "{Active red giants: Close binaries versus single rapid rotators}",
      journal = {\aap},
         year = 2020,
        month = jul,
       volume = {639},
          eid = {A63},
        pages = {A63},
          doi = {10.1051/0004-6361/202037781},
archivePrefix = {arXiv},
       eprint = {2004.13792},
 primaryClass = {astro-ph.SR},
       adsurl = {https://ui.adsabs.harvard.edu/abs/2020A&A...639A..63G}
}

@ARTICLE{huber+09,
       author = {{Huber}, D. and {Stello}, D. and {Bedding}, T.~R. and {Chaplin}, W.~J. and {Arentoft}, T. and {Quirion}, P.-O. and {Kjeldsen}, H.},
        title = "{Automated extraction of oscillation parameters for Kepler observations of solar-type stars}",
      journal = {Communications in Asteroseismology},
         year = 2009,
        month = oct,
       volume = {160},
        pages = {74},
          doi = {10.48550/arXiv.0910.2764},
archivePrefix = {arXiv},
       eprint = {0910.2764},
 primaryClass = {astro-ph.SR},
       adsurl = {https://ui.adsabs.harvard.edu/abs/2009CoAst.160...74H}
}

@ARTICLE{hekker+10,
       author = {{Hekker}, S. and {Broomhall}, A.-M. and {Chaplin}, W.~J. and {Elsworth}, Y.~P. and {Fletcher}, S.~T. and {New}, R. and {Arentoft}, T. and {Quirion}, P.-O. and {Kjeldsen}, H.},
        title = "{The Octave (Birmingham-Sheffield Hallam) automated pipeline for extracting oscillation parameters of solar-like main-sequence stars}",
      journal = {\mnras},
         year = 2010,
        month = mar,
       volume = {402},
       number = {3},
        pages = {2049-2059},
          doi = {10.1111/j.1365-2966.2009.16030.x},
archivePrefix = {arXiv},
       eprint = {0911.2612},
 primaryClass = {astro-ph.SR},
       adsurl = {https://ui.adsabs.harvard.edu/abs/2010MNRAS.402.2049H}
}

@ARTICLE{kallinger+10,
       author = {{Kallinger}, T. and {Mosser}, B. and {Hekker}, S. and {Huber}, D. and {Stello}, D. and {Mathur}, S. and {Basu}, S. and {Bedding}, T.~R. and {Chaplin}, W.~J. and {De Ridder}, J. and {Elsworth}, Y.~P. and {Frandsen}, S. and {Garc{\'\i}a}, R.~A. and {Gruberbauer}, M. and {Matthews}, J.~M. and {Borucki}, W.~J. and {Bruntt}, H. and {Christensen-Dalsgaard}, J. and {Gilliland}, R.~L. and {Kjeldsen}, H. and {Koch}, D.~G.},
        title = "{Asteroseismology of red giants from the first four months of Kepler data: Fundamental stellar parameters}",
      journal = {\aap},
         year = 2010,
        month = nov,
       volume = {522},
          eid = {A1},
        pages = {A1},
          doi = {10.1051/0004-6361/201015263},
archivePrefix = {arXiv},
       eprint = {1010.4589},
 primaryClass = {astro-ph.SR},
       adsurl = {https://ui.adsabs.harvard.edu/abs/2010A&A...522A...1K}
}

@ARTICLE{bedding+10,
       author = {{Bedding}, T.~R. and {Huber}, D. and {Stello}, D. and {Elsworth}, Y.~P. and {Hekker}, S. and {Kallinger}, T. and {Mathur}, S. and {Mosser}, B. and {Preston}, H.~L. and {Ballot}, J. and {Barban}, C. and {Broomhall}, A.~M. and {Buzasi}, D.~L. and {Chaplin}, W.~J. and {Garc{\'\i}a}, R.~A. and {Gruberbauer}, M. and {Hale}, S.~J. and {De Ridder}, J. and {Frandsen}, S. and {Borucki}, W.~J. and {Brown}, T. and {Christensen-Dalsgaard}, J. and {Gilliland}, R.~L. and {Jenkins}, J.~M. and {Kjeldsen}, H. and {Koch}, D. and {Belkacem}, K. and {Bildsten}, L. and {Bruntt}, H. and {Campante}, T.~L. and {Deheuvels}, S. and {Derekas}, A. and {Dupret}, M.-A. and {Goupil}, M.-J. and {Hatzes}, A. and {Houdek}, G. and {Ireland}, M.~J. and {Jiang}, C. and {Karoff}, C. and {Kiss}, L.~L. and {Lebreton}, Y. and {Miglio}, A. and {Montalb{\'a}n}, J. and {Noels}, A. and {Roxburgh}, I.~W. and {Sangaralingam}, V. and {Stevens}, I.~R. and {Suran}, M.~D. and {Tarrant}, N.~J. and {Weiss}, A.},
        title = "{Solar-like Oscillations in Low-luminosity Red Giants: First Results from Kepler}",
      journal = {\apjl},
         year = 2010,
        month = apr,
       volume = {713},
       number = {2},
        pages = {L176-L181},
          doi = {10.1088/2041-8205/713/2/L176},
archivePrefix = {arXiv},
       eprint = {1001.0229},
 primaryClass = {astro-ph.SR},
       adsurl = {https://ui.adsabs.harvard.edu/abs/2010ApJ...713L.176B}
}

@ARTICLE{stello+09,
       author = {{Stello}, Dennis and {Chaplin}, William J. and {Bruntt}, Hans and {Creevey}, Orlagh L. and {Garc{\'\i}a-Hern{\'a}ndez}, Antonio and {Monteiro}, Mario J.~P.~F.~G. and {Moya}, Andr{\'e}s and {Quirion}, Pierre-Olivier and {Sousa}, Sergio G. and {Su{\'a}rez}, Juan-Carlos and {Appourchaux}, Thierry and {Arentoft}, Torben and {Ballot}, Jerome and {Bedding}, Timothy R. and {Christensen-Dalsgaard}, J{\o}rgen and {Elsworth}, Yvonne and {Fletcher}, Stephen T. and {Garc{\'\i}a}, Rafael A. and {Houdek}, G{\"u}nter and {Jim{\'e}nez-Reyes}, Sebastian J. and {Kjeldsen}, Hans and {New}, Roger and {R{\'e}gulo}, Clara and {Salabert}, David and {Toutain}, Thierry},
        title = "{Radius Determination of Solar-type Stars Using Asteroseismology: What to Expect from the Kepler Mission}",
      journal = {\apj},
         year = 2009,
        month = aug,
       volume = {700},
       number = {2},
        pages = {1589-1602},
          doi = {10.1088/0004-637X/700/2/1589},
archivePrefix = {arXiv},
       eprint = {0906.0766},
 primaryClass = {astro-ph.SR},
       adsurl = {https://ui.adsabs.harvard.edu/abs/2009ApJ...700.1589S}
}

@ARTICLE{appourchaux+08,
       author = {{Appourchaux}, T. and {Michel}, E. and {Auvergne}, M. and {Baglin}, A. and {Toutain}, T. and {Baudin}, F. and {Benomar}, O. and {Chaplin}, W.~J. and {Deheuvels}, S. and {Samadi}, R. and {Verner}, G.~A. and {Boumier}, P. and {Garc{\'\i}a}, R.~A. and {Mosser}, B. and {Hulot}, J.-C. and {Ballot}, J. and {Barban}, C. and {Elsworth}, Y. and {Jim{\'e}nez-Reyes}, S.~J. and {Kjeldsen}, H. and {R{\'e}gulo}, C. and {Roxburgh}, I.~W.},
        title = "{CoRoT sounds the stars: p-mode parameters of Sun-like oscillations on HD 49933}",
      journal = {\aap},
         year = 2008,
        month = sep,
       volume = {488},
       number = {2},
        pages = {705-714},
          doi = {10.1051/0004-6361:200810297},
       adsurl = {https://ui.adsabs.harvard.edu/abs/2008A&A...488..705A}
}

@ARTICLE{khan+18,
       author = {{Khan}, Saniya and {Hall}, Oliver J. and {Miglio}, Andrea and {Davies}, Guy R. and {Mosser}, Beno{\^\i}t and {Girardi}, L{\'e}o and {Montalb{\'a}n}, Josefina},
        title = "{The Red-giant Branch Bump Revisited: Constraints on Envelope Overshooting in a Wide Range of Masses and Metallicities}",
      journal = {\apj},
         year = 2018,
        month = jun,
       volume = {859},
       number = {2},
          eid = {156},
        pages = {156},
          doi = {10.3847/1538-4357/aabf90},
archivePrefix = {arXiv},
       eprint = {1804.06669},
 primaryClass = {astro-ph.SR},
       adsurl = {https://ui.adsabs.harvard.edu/abs/2018ApJ...859..156K}
}

@ARTICLE{buchele25,
       author = {{Buchele}, Lynn},
        title = "{Exploring Mixing Thresholds in Asteroseismic Stellar Evolution Models}",
      journal = {Research Notes of the American Astronomical Society},
         year = 2025,
        month = jul,
       volume = {9},
       number = {7},
          eid = {193},
        pages = {193},
          doi = {10.3847/2515-5172/adf06d},
archivePrefix = {arXiv},
       eprint = {2507.23275},
 primaryClass = {astro-ph.SR},
       adsurl = {https://ui.adsabs.harvard.edu/abs/2025RNAAS...9..193B}
}

@ARTICLE{garcia+ballot19,
       author = {{Garc{\'\i}a}, Rafael A. and {Ballot}, J{\'e}r{\^o}me},
        title = "{Asteroseismology of solar-type stars}",
      journal = {Living Reviews in Solar Physics},
         year = 2019,
        month = dec,
       volume = {16},
       number = {1},
          eid = {4},
        pages = {4},
          doi = {10.1007/s41116-019-0020-1},
archivePrefix = {arXiv},
       eprint = {1906.12262},
 primaryClass = {astro-ph.SR},
       adsurl = {https://ui.adsabs.harvard.edu/abs/2019LRSP...16....4G}
}

@ARTICLE{chaplin+miglio13,
       author = {{Chaplin}, William J. and {Miglio}, Andrea},
        title = "{Asteroseismology of Solar-Type and Red-Giant Stars}",
      journal = {\araa},
         year = 2013,
        month = aug,
       volume = {51},
       number = {1},
        pages = {353-392},
          doi = {10.1146/annurev-astro-082812-140938},
archivePrefix = {arXiv},
       eprint = {1303.1957},
 primaryClass = {astro-ph.SR},
       adsurl = {https://ui.adsabs.harvard.edu/abs/2013ARA&A..51..353C}
}

@ARTICLE{deRidder+09,
       author = {{De Ridder}, Joris and {Barban}, Caroline and {Baudin}, Fr{\'e}d{\'e}ric and {Carrier}, Fabien and {Hatzes}, Artie P. and {Hekker}, Saskia and {Kallinger}, Thomas and {Weiss}, Werner W. and {Baglin}, Annie and {Auvergne}, Michel and {Samadi}, R{\'e}za and {Barge}, Pierre and {Deleuil}, Magali},
        title = "{Non-radial oscillation modes with long lifetimes in giant stars}",
      journal = {\nat},
         year = 2009,
        month = may,
       volume = {459},
       number = {7245},
        pages = {398-400},
          doi = {10.1038/nature08022},
       adsurl = {https://ui.adsabs.harvard.edu/abs/2009Natur.459..398D}
}

@ARTICLE{vanLier+25,
       author = {{van Lier}, T. and {M{\"u}ller}, J. and {Hekker}, S.},
        title = "{Weak or strong: Coupling of mixed oscillation modes on the red giant branch}",
      journal = {\aap},
         year = 2025,
        month = aug,
       volume = {700},
          eid = {A1},
        pages = {A1},
          doi = {10.1051/0004-6361/202555255},
archivePrefix = {arXiv},
       eprint = {2506.04829},
 primaryClass = {astro-ph.SR},
       adsurl = {https://ui.adsabs.harvard.edu/abs/2025A&A...700A...1V}
}

@ARTICLE{david+26,
       author = {{David}, Cy S. and {Lecoanet}, Daniel and {Garaud}, Pascale},
        title = "{Conversion and Damping of Nonaxisymmetric Internal Gravity Waves in Magnetized Stellar Cores}",
      journal = {\apj},
         year = 2026,
        month = apr,
       volume = {1000},
       number = {2},
          eid = {292},
        pages = {292},
          doi = {10.3847/1538-4357/ae4d18},
archivePrefix = {arXiv},
       eprint = {2510.14026},
 primaryClass = {astro-ph.SR},
       adsurl = {https://ui.adsabs.harvard.edu/abs/2026ApJ..1000..292D}
}

@ARTICLE{vanHoolst+98,
       author = {{Van Hoolst}, T. and {Dziembowski}, W.~A. and {Kawaler}, S.~D.},
        title = "{Unstable non-radial modes in radial pulsators: theory and an example}",
      journal = {\mnras},
         year = 1998,
        month = jun,
       volume = {297},
       number = {2},
        pages = {536-544},
          doi = {10.1046/j.1365-8711.1998.01540.x},
       adsurl = {https://ui.adsabs.harvard.edu/abs/1998MNRAS.297..536V}
}

@ARTICLE{dziembowski77,
       author = {{Dziembowski}, W.},
        title = "{Oscillations of giants and supergiants.}",
      journal = {\actaa},
         year = 1977,
        month = jan,
       volume = {27},
        pages = {95-126},
       adsurl = {https://ui.adsabs.harvard.edu/abs/1977AcA....27...95D}
}

@ARTICLE{corsaro+15,
       author = {{Corsaro}, E. and {De Ridder}, J. and {Garc{\'\i}a}, R.~A.},
        title = "{Bayesian peak bagging analysis of 19 low-mass low-luminosity red giants observed with Kepler}",
      journal = {\aap},
         year = 2015,
        month = jul,
       volume = {579},
          eid = {A83},
        pages = {A83},
          doi = {10.1051/0004-6361/201525895},
archivePrefix = {arXiv},
       eprint = {1503.08821},
 primaryClass = {astro-ph.SR},
       adsurl = {https://ui.adsabs.harvard.edu/abs/2015A&A...579A..83C}
}

@ARTICLE{baudin+11,
       author = {{Baudin}, F. and {Barban}, C. and {Belkacem}, K. and {Hekker}, S. and {Morel}, T. and {Samadi}, R. and {Benomar}, O. and {Goupil}, M.-J. and {Carrier}, F. and {Ballot}, J. and {Deheuvels}, S. and {De Ridder}, J. and {Hatzes}, A.~P. and {Kallinger}, T. and {Weiss}, W.~W.},
        title = "{Amplitudes and lifetimes of solar-like oscillations observed by CoRoT. Red-giant versus main-sequence stars}",
      journal = {\aap},
         year = 2011,
        month = may,
       volume = {529},
          eid = {A84},
        pages = {A84},
          doi = {10.1051/0004-6361/201014037},
archivePrefix = {arXiv},
       eprint = {1102.1896},
 primaryClass = {astro-ph.SR},
       adsurl = {https://ui.adsabs.harvard.edu/abs/2011A&A...529A..84B}
}

@ARTICLE{vrard+18,
       author = {{Vrard}, M. and {Kallinger}, T. and {Mosser}, B. and {Barban}, C. and {Baudin}, F. and {Belkacem}, K. and {Cunha}, M.~S.},
        title = "{Amplitude and lifetime of radial modes in red giant star spectra observed by Kepler}",
      journal = {\aap},
         year = 2018,
        month = aug,
       volume = {616},
          eid = {A94},
        pages = {A94},
          doi = {10.1051/0004-6361/201732477},
archivePrefix = {arXiv},
       eprint = {1805.03690},
 primaryClass = {astro-ph.SR},
       adsurl = {https://ui.adsabs.harvard.edu/abs/2018A&A...616A..94V}
}

@INPROCEEDINGS{samadi+15,
       author = {{Samadi}, R. and {Belkacem}, K. and {Sonoi}, T.},
        title = "{Stellar oscillations - II - The non-adiabatic case}",
    booktitle = {EAS Publications Series},
         year = 2015,
       series = {EAS Publications Series},
       volume = {73-74},
        month = feb,
    publisher = {EDP},
        pages = {111-191},
          doi = {10.1051/eas/1573003},
archivePrefix = {arXiv},
       eprint = {1510.01151},
 primaryClass = {astro-ph.SR},
       adsurl = {https://ui.adsabs.harvard.edu/abs/2015EAS....73..111S}
}

@ARTICLE{xiong+97,
       author = {{Xiong}, D.~R. and {Cheng}, Q.~L. and {Deng}, L.},
        title = "{Nonlocal Time-dependent Convection Theory}",
      journal = {\apjs},
         year = 1997,
        month = feb,
       volume = {108},
       number = {2},
        pages = {529-544},
          doi = {10.1086/312959},
       adsurl = {https://ui.adsabs.harvard.edu/abs/1997ApJS..108..529X}
}

@ARTICLE{gough77,
       author = {{Gough}, D.~O.},
        title = "{Mixing-length theory for pulsating stars.}",
      journal = {\apj},
         year = 1977,
        month = may,
       volume = {214},
        pages = {196-213},
          doi = {10.1086/155244},
       adsurl = {https://ui.adsabs.harvard.edu/abs/1977ApJ...214..196G}
}

@ARTICLE{unno67,
       author = {{Unno}, Wasaburo},
        title = "{The Stellar Radial Pulsation Coupled with the Convection}",
      journal = {\pasj},
         year = 1967,
        month = jun,
       volume = {19},
       number = {2},
        pages = {140-153},
          doi = {10.1093/pasj/19.2.140},
       adsurl = {https://ui.adsabs.harvard.edu/abs/1967PASJ...19..140U}
}

@ARTICLE{grigahcene05,
       author = {{Grigahc{\`e}ne}, A. and {Dupret}, M.-A. and {Gabriel}, M. and {Garrido}, R. and {Scuflaire}, R.},
        title = "{Convection-pulsation coupling. I. A mixing-length perturbative theory}",
      journal = {\aap},
         year = 2005,
        month = may,
       volume = {434},
       number = {3},
        pages = {1055-1062},
          doi = {10.1051/0004-6361:20041816},
       adsurl = {https://ui.adsabs.harvard.edu/abs/2005A&A...434.1055G}
}

@ARTICLE{gabriel96,
       author = {{Gabriel}, M.},
        title = "{Solar oscillations : theory}",
      journal = {Bulletin of the Astronomical Society of India},
         year = 1996,
        month = jun,
       volume = {24},
        pages = {233},
       adsurl = {https://ui.adsabs.harvard.edu/abs/1996BASI...24..233G}
}

@ARTICLE{dziembowski+01,
       author = {{Dziembowski}, W.~A. and {Gough}, D.~O. and {Houdek}, G. and {Sienkiewicz}, R.},
        title = "{Oscillations of {\ensuremath{\alpha}} UMa and other red giants}",
      journal = {\mnras},
         year = 2001,
        month = dec,
       volume = {328},
       number = {2},
        pages = {601-610},
          doi = {10.1046/j.1365-8711.2001.04894.x},
archivePrefix = {arXiv},
       eprint = {astro-ph/0108337},
 primaryClass = {astro-ph},
       adsurl = {https://ui.adsabs.harvard.edu/abs/2001MNRAS.328..601D}
}

@ARTICLE{muller+26,
       author = {{M{\"u}ller}, Jonas and {Copp{\'e}e}, Quentin and {Van Beeck}, Jordan and {van Lier}, Tobias and {Hekker}, Saskia},
        title = "{Asymptotic power spectra and visibilities of damped mixed modes}",
      journal = {\aap},
         year = 2026,
        month = mar,
       volume = {707},
          eid = {A126},
        pages = {A126},
          doi = {10.1051/0004-6361/202557877},
archivePrefix = {arXiv},
       eprint = {2601.21745},
 primaryClass = {astro-ph.SR},
       adsurl = {https://ui.adsabs.harvard.edu/abs/2026A&A...707A.126M}
}

@ARTICLE{cunha+15,
       author = {{Cunha}, M.~S. and {Stello}, D. and {Avelino}, P.~P. and {Christensen-Dalsgaard}, J. and {Townsend}, R.~H.~D.},
        title = "{Structural Glitches near the Cores of Red Giants Revealed by Oscillations in g-mode Period Spacings from Stellar Models}",
      journal = {\apj},
         year = 2015,
        month = jun,
       volume = {805},
       number = {2},
          eid = {127},
        pages = {127},
          doi = {10.1088/0004-637X/805/2/127},
archivePrefix = {arXiv},
       eprint = {1503.09085},
 primaryClass = {astro-ph.SR},
       adsurl = {https://ui.adsabs.harvard.edu/abs/2015ApJ...805..127C}
}

@ARTICLE{vanRossem+24,
       author = {{van Rossem}, Walter E. and {Miglio}, Andrea and {Montalb{\'a}n}, Josefina},
        title = "{Mixed-mode coupling in the red clump: I. Standard single star models}",
      journal = {\aap},
         year = 2024,
        month = nov,
       volume = {691},
          eid = {A177},
        pages = {A177},
          doi = {10.1051/0004-6361/202451281},
archivePrefix = {arXiv},
       eprint = {2410.02865},
 primaryClass = {astro-ph.SR},
       adsurl = {https://ui.adsabs.harvard.edu/abs/2024A&A...691A.177V}
}

@ARTICLE{cowling41,
       author = {{Cowling}, T.~G.},
        title = "{The non-radial oscillations of polytropic stars}",
      journal = {\mnras},
         year = 1941,
        month = jan,
       volume = {101},
        pages = {367},
          doi = {10.1093/mnras/101.8.367},
       adsurl = {https://ui.adsabs.harvard.edu/abs/1941MNRAS.101..367C}
}

@ARTICLE{dreau+21,
       author = {{Dr{\'e}au}, G. and {Mosser}, B. and {Lebreton}, Y. and {Gehan}, C. and {Kallinger}, T.},
        title = "{Seismic constraints on the internal structure of evolved stars: From high-luminosity RGB to AGB stars}",
      journal = {\aap},
         year = 2021,
        month = jun,
       volume = {650},
          eid = {A115},
        pages = {A115},
          doi = {10.1051/0004-6361/202040240},
archivePrefix = {arXiv},
       eprint = {2103.16718},
 primaryClass = {astro-ph.SR},
       adsurl = {https://ui.adsabs.harvard.edu/abs/2021A&A...650A.115D}
}

@ARTICLE{mosser+17coupling,
       author = {{Mosser}, B. and {Pin{\c{c}}on}, C. and {Belkacem}, K. and {Takata}, M. and {Vrard}, M.},
        title = "{Period spacings in red giants. III. Coupling factors of mixed modes}",
      journal = {\aap},
         year = 2017,
        month = apr,
       volume = {600},
          eid = {A1},
        pages = {A1},
          doi = {10.1051/0004-6361/201630053},
archivePrefix = {arXiv},
       eprint = {1612.08453},
 primaryClass = {astro-ph.SR},
       adsurl = {https://ui.adsabs.harvard.edu/abs/2017A&A...600A...1M}
}

@ARTICLE{jiang+22,
       author = {{Jiang}, C. and {Cunha}, M. and {Christensen-Dalsgaard}, J. and {Zhang}, Q.~S. and {Gizon}, L.},
        title = "{Evolution of dipolar mixed-mode coupling factor in red giant stars: impact of buoyancy spike}",
      journal = {\mnras},
         year = 2022,
        month = sep,
       volume = {515},
       number = {3},
        pages = {3853-3866},
          doi = {10.1093/mnras/stac2065},
archivePrefix = {arXiv},
       eprint = {2207.09878},
 primaryClass = {astro-ph.SR},
       adsurl = {https://ui.adsabs.harvard.edu/abs/2022MNRAS.515.3853J}
}

@ARTICLE{mesa6,
       author = {{Jermyn}, Adam S. and {Bauer}, Evan B. and {Schwab}, Josiah and {Farmer}, R. and {Ball}, Warrick H. and {Bellinger}, Earl P. and {Dotter}, Aaron and {Joyce}, Meridith and {Marchant}, Pablo and {Mombarg}, Joey S.~G. and {Wolf}, William M. and {Sunny Wong}, Tin Long and {Cinquegrana}, Giulia C. and {Farrell}, Eoin and {Smolec}, R. and {Thoul}, Anne and {Cantiello}, Matteo and {Herwig}, Falk and {Toloza}, Odette and {Bildsten}, Lars and {Townsend}, Richard H.~D. and {Timmes}, F.~X.},
        title = "{Modules for Experiments in Stellar Astrophysics (MESA): Time-dependent Convection, Energy Conservation, Automatic Differentiation, and Infrastructure}",
      journal = {\apjs},
         year = 2023,
        month = mar,
       volume = {265},
       number = {1},
          eid = {15},
        pages = {15},
          doi = {10.3847/1538-4365/acae8d},
archivePrefix = {arXiv},
       eprint = {2208.03651},
 primaryClass = {astro-ph.SR},
       adsurl = {https://ui.adsabs.harvard.edu/abs/2023ApJS..265...15J}
}

@ARTICLE{muller+25,
       author = {{M{\"u}ller}, Jonas and {Copp{\'e}e}, Quentin and {Hekker}, Saskia},
        title = "{Oscillations of red giant stars with magnetic damping in the core: I. Dissipation of mode energy in dipole-like magnetic fields}",
      journal = {\aap},
         year = 2025,
        month = apr,
       volume = {696},
          eid = {A134},
        pages = {A134},
          doi = {10.1051/0004-6361/202553888},
archivePrefix = {arXiv},
       eprint = {2503.11451},
 primaryClass = {astro-ph.SR},
       adsurl = {https://ui.adsabs.harvard.edu/abs/2025A&A...696A.134M}
}

@ARTICLE{coppee+24,
       author = {{Copp{\'e}e}, Q. and {M{\"u}ller}, J. and {Bazot}, M. and {Hekker}, S.},
        title = "{The radial modes of stars with suppressed dipole modes}",
      journal = {\aap},
         year = 2024,
        month = oct,
       volume = {690},
          eid = {A324},
        pages = {A324},
          doi = {10.1051/0004-6361/202450037},
archivePrefix = {arXiv},
       eprint = {2409.12692},
 primaryClass = {astro-ph.SR},
       adsurl = {https://ui.adsabs.harvard.edu/abs/2024A&A...690A.324C}
}

@ARTICLE{garcia+14,
       author = {{Garc{\'\i}a}, R.~A. and {P{\'e}rez Hern{\'a}ndez}, F. and {Benomar}, O. and {Silva Aguirre}, V. and {Ballot}, J. and {Davies}, G.~R. and {Do{\u{g}}an}, G. and {Stello}, D. and {Christensen-Dalsgaard}, J. and {Houdek}, G. and {Ligni{\`e}res}, F. and {Mathur}, S. and {Takata}, M. and {Ceillier}, T. and {Chaplin}, W.~J. and {Mathis}, S. and {Mosser}, B. and {Ouazzani}, R.~M. and {Pinsonneault}, M.~H. and {Reese}, D.~R. and {R{\'e}gulo}, C. and {Salabert}, D. and {Thompson}, M.~J. and {van Saders}, J.~L. and {Neiner}, C. and {De Ridder}, J.},
        title = "{Study of KIC 8561221 observed by Kepler: an early red giant showing depressed dipolar modes}",
      journal = {\aap},
         year = 2014,
        month = mar,
       volume = {563},
          eid = {A84},
        pages = {A84},
          doi = {10.1051/0004-6361/201322823},
archivePrefix = {arXiv},
       eprint = {1311.6990},
 primaryClass = {astro-ph.SR},
       adsurl = {https://ui.adsabs.harvard.edu/abs/2014A&A...563A..84G}
}

@ARTICLE{takata05,
       author = {{Takata}, Masao},
        title = "{Momentum Conservation and Mode Classification of the Dipolar Oscillations of Stars}",
      journal = {\pasj},
         year = 2005,
        month = apr,
       volume = {57},
        pages = {375-389},
          doi = {10.1093/pasj/57.2.375},
       adsurl = {https://ui.adsabs.harvard.edu/abs/2005PASJ...57..375T}
}

@ARTICLE{pincon+19,
       author = {{Pin{\c{c}}on}, C. and {Takata}, M. and {Mosser}, B.},
        title = "{Evolution of the gravity offset of mixed modes in RGB stars}",
      journal = {\aap},
         year = 2019,
        month = jun,
       volume = {626},
          eid = {A125},
        pages = {A125},
          doi = {10.1051/0004-6361/201935327},
archivePrefix = {arXiv},
       eprint = {1905.05691},
 primaryClass = {astro-ph.SR},
       adsurl = {https://ui.adsabs.harvard.edu/abs/2019A&A...626A.125P}
}

@INCOLLECTION{gough90,
       author = {{Gough}, D.~O.},
        title = "{Comments on Helioseismic Inference}",
    booktitle = {Progress of Seismology of the Sun and Stars},
         year = 1990,
       editor = {{Osaki}, Yoji and {Shibahashi}, Hiromoto},
       volume = {367},
        pages = {283},
          doi = {10.1007/3-540-53091-610.1007/3-540-53091-6_93},
       adsurl = {https://ui.adsabs.harvard.edu/abs/1990LNP...367..283G}
}

@ARTICLE{takata06a,
       author = {{Takata}, Masao},
        title = "{Analysis of Adiabatic Dipolar Oscillations of Stars}",
      journal = {\pasj},
         year = 2006,
        month = oct,
       volume = {58},
        pages = {893-908},
          doi = {10.1093/pasj/58.5.893},
       adsurl = {https://ui.adsabs.harvard.edu/abs/2006PASJ...58..893T}
}

@BOOK{aerts+10book,
       author = {{Aerts}, Conny and {Christensen-Dalsgaard}, J{\o}rgen and {Kurtz}, Donald W.},
        title = "{Asteroseismology}",
         year = 2010,
          doi = {10.1007/978-1-4020-5803-5},
       adsurl = {https://ui.adsabs.harvard.edu/abs/2010aste.book.....A}
}

@ARTICLE{pincon+20,
       author = {{Pin{\c{c}}on}, C. and {Goupil}, M.~J. and {Belkacem}, K.},
        title = "{Probing the mid-layer structure of red giants. I. Mixed-mode coupling factor as a seismic diagnosis}",
      journal = {\aap},
         year = 2020,
        month = feb,
       volume = {634},
          eid = {A68},
        pages = {A68},
          doi = {10.1051/0004-6361/201936864},
archivePrefix = {arXiv},
       eprint = {1912.06008},
 primaryClass = {astro-ph.SR},
       adsurl = {https://ui.adsabs.harvard.edu/abs/2020A&A...634A..68P}
}

@ARTICLE{hatt+24,
       author = {{Hatt}, Emily J. and {Ong}, J.~M. Joel and {Nielsen}, Martin B. and {Chaplin}, William J. and {Davies}, Guy R. and {Deheuvels}, S{\'e}bastien and {Ballot}, J{\'e}r{\^o}me and {Li}, Gang and {Bugnet}, Lisa},
        title = "{Asteroseismic signatures of core magnetism and rotation in hundreds of low-luminosity red giants}",
      journal = {\mnras},
         year = 2024,
        month = oct,
       volume = {534},
       number = {2},
        pages = {1060-1076},
          doi = {10.1093/mnras/stae2053},
archivePrefix = {arXiv},
       eprint = {2409.01157},
 primaryClass = {astro-ph.SR},
       adsurl = {https://ui.adsabs.harvard.edu/abs/2024MNRAS.534.1060H}
}

@ARTICLE{corsaro+12,
       author = {{Corsaro}, Enrico and {Stello}, Dennis and {Huber}, Daniel and {Bedding}, Timothy R. and {Bonanno}, Alfio and {Brogaard}, Karsten and {Kallinger}, Thomas and {Benomar}, Othman and {White}, Timothy R. and {Mosser}, Benoit and {Basu}, Sarbani and {Chaplin}, William J. and {Christensen-Dalsgaard}, J{\o}rgen and {Elsworth}, Yvonne P. and {Garc{\'\i}a}, Rafael A. and {Hekker}, Saskia and {Kjeldsen}, Hans and {Mathur}, Savita and {Meibom}, S{\o}ren and {Hall}, Jennifer R. and {Ibrahim}, Khadeejah A. and {Klaus}, Todd C.},
        title = "{Asteroseismology of the Open Clusters NGC 6791, NGC 6811, and NGC 6819 from 19 Months of Kepler Photometry}",
      journal = {\apj},
         year = 2012,
        month = oct,
       volume = {757},
       number = {2},
          eid = {190},
        pages = {190},
          doi = {10.1088/0004-637X/757/2/190},
archivePrefix = {arXiv},
       eprint = {1205.4023},
 primaryClass = {astro-ph.SR},
       adsurl = {https://ui.adsabs.harvard.edu/abs/2012ApJ...757..190C}
}

@ARTICLE{shibahashi79,
       author = {{Shibahashi}, H.},
        title = "{Modal Analysis of Stellar Nonradial Oscillations by an Asymptotic Method}",
      journal = {\pasj},
         year = 1979,
        month = jan,
       volume = {31},
        pages = {87-104},
       adsurl = {https://ui.adsabs.harvard.edu/abs/1979PASJ...31...87S}
}

@BOOK{unno+89,
       author = {{Unno}, Wasaburo and {Osaki}, Yoji and {Ando}, Hiroyasu and {Saio}, H. and {Shibahashi}, H.},
        title = "{Nonradial oscillations of stars}",
         year = 1989,
       adsurl = {https://ui.adsabs.harvard.edu/abs/1989nos..book.....U}
}

@ARTICLE{mesa5,
       author = {{Paxton}, Bill and {Smolec}, R. and {Schwab}, Josiah and {Gautschy}, A. and {Bildsten}, Lars and {Cantiello}, Matteo and {Dotter}, Aaron and {Farmer}, R. and {Goldberg}, Jared A. and {Jermyn}, Adam S. and {Kanbur}, S.~M. and {Marchant}, Pablo and {Thoul}, Anne and {Townsend}, Richard H.~D. and {Wolf}, William M. and {Zhang}, Michael and {Timmes}, F.~X.},
        title = "{Modules for Experiments in Stellar Astrophysics (MESA): Pulsating Variable Stars, Rotation, Convective Boundaries, and Energy Conservation}",
      journal = {\apjs},
         year = 2019,
        month = jul,
       volume = {243},
       number = {1},
          eid = {10},
        pages = {10},
          doi = {10.3847/1538-4365/ab2241},
archivePrefix = {arXiv},
       eprint = {1903.01426},
 primaryClass = {astro-ph.SR},
       adsurl = {https://ui.adsabs.harvard.edu/abs/2019ApJS..243...10P}
}

@ARTICLE{mesa4,
       author = {{Paxton}, Bill and {Schwab}, Josiah and {Bauer}, Evan B. and {Bildsten}, Lars and {Blinnikov}, Sergei and {Duffell}, Paul and {Farmer}, R. and {Goldberg}, Jared A. and {Marchant}, Pablo and {Sorokina}, Elena and {Thoul}, Anne and {Townsend}, Richard H.~D. and {Timmes}, F.~X.},
        title = "{Modules for Experiments in Stellar Astrophysics (MESA): Convective Boundaries, Element Diffusion, and Massive Star Explosions}",
      journal = {\apjs},
         year = 2018,
        month = feb,
       volume = {234},
       number = {2},
          eid = {34},
        pages = {34},
          doi = {10.3847/1538-4365/aaa5a8},
archivePrefix = {arXiv},
       eprint = {1710.08424},
 primaryClass = {astro-ph.SR},
       adsurl = {https://ui.adsabs.harvard.edu/abs/2018ApJS..234...34P}
}

@ARTICLE{mesa3,
       author = {{Paxton}, Bill and {Marchant}, Pablo and {Schwab}, Josiah and {Bauer}, Evan B. and {Bildsten}, Lars and {Cantiello}, Matteo and {Dessart}, Luc and {Farmer}, R. and {Hu}, H. and {Langer}, N. and {Townsend}, R.~H.~D. and {Townsley}, Dean M. and {Timmes}, F.~X.},
        title = "{Modules for Experiments in Stellar Astrophysics (MESA): Binaries, Pulsations, and Explosions}",
      journal = {\apjs},
         year = 2015,
        month = sep,
       volume = {220},
       number = {1},
          eid = {15},
        pages = {15},
          doi = {10.1088/0067-0049/220/1/15},
archivePrefix = {arXiv},
       eprint = {1506.03146},
 primaryClass = {astro-ph.SR},
       adsurl = {https://ui.adsabs.harvard.edu/abs/2015ApJS..220...15P}
}

@ARTICLE{mesa2,
       author = {{Paxton}, Bill and {Cantiello}, Matteo and {Arras}, Phil and {Bildsten}, Lars and {Brown}, Edward F. and {Dotter}, Aaron and {Mankovich}, Christopher and {Montgomery}, M.~H. and {Stello}, Dennis and {Timmes}, F.~X. and {Townsend}, Richard},
        title = "{Modules for Experiments in Stellar Astrophysics (MESA): Planets, Oscillations, Rotation, and Massive Stars}",
      journal = {\apjs},
         year = 2013,
        month = sep,
       volume = {208},
       number = {1},
          eid = {4},
        pages = {4},
          doi = {10.1088/0067-0049/208/1/4},
archivePrefix = {arXiv},
       eprint = {1301.0319},
 primaryClass = {astro-ph.SR},
       adsurl = {https://ui.adsabs.harvard.edu/abs/2013ApJS..208....4P}
}

@ARTICLE{mesa1,
       author = {{Paxton}, Bill and {Bildsten}, Lars and {Dotter}, Aaron and {Herwig}, Falk and {Lesaffre}, Pierre and {Timmes}, Frank},
        title = "{Modules for Experiments in Stellar Astrophysics (MESA)}",
      journal = {\apjs},
         year = 2011,
        month = jan,
       volume = {192},
       number = {1},
          eid = {3},
        pages = {3},
          doi = {10.1088/0067-0049/192/1/3},
archivePrefix = {arXiv},
       eprint = {1009.1622},
 primaryClass = {astro-ph.SR},
       adsurl = {https://ui.adsabs.harvard.edu/abs/2011ApJS..192....3P}
}

@ARTICLE{rui+fuller23,
       author = {{Rui}, Nicholas Z. and {Fuller}, Jim},
        title = "{Gravity waves in strong magnetic fields}",
      journal = {\mnras},
         year = 2023,
        month = jul,
       volume = {523},
       number = {1},
        pages = {582-602},
          doi = {10.1093/mnras/stad1424},
archivePrefix = {arXiv},
       eprint = {2303.08147},
 primaryClass = {astro-ph.SR},
       adsurl = {https://ui.adsabs.harvard.edu/abs/2023MNRAS.523..582R}
}

@ARTICLE{lecoanet+17,
       author = {{Lecoanet}, D. and {Vasil}, G.~M. and {Fuller}, J. and {Cantiello}, M. and {Burns}, K.~J.},
        title = "{Conversion of internal gravity waves into magnetic waves}",
      journal = {\mnras},
         year = 2017,
        month = apr,
       volume = {466},
       number = {2},
        pages = {2181-2193},
          doi = {10.1093/mnras/stw3273},
archivePrefix = {arXiv},
       eprint = {1610.08506},
 primaryClass = {astro-ph.SR},
       adsurl = {https://ui.adsabs.harvard.edu/abs/2017MNRAS.466.2181L}
}

@ARTICLE{stello+16b,
       author = {{Stello}, Dennis and {Cantiello}, Matteo and {Fuller}, Jim and {Huber}, Daniel and {Garc{\'\i}a}, Rafael A. and {Bedding}, Timothy R. and {Bildsten}, Lars and {Silva Aguirre}, Victor},
        title = "{A prevalence of dynamo-generated magnetic fields in the cores of intermediate-mass stars}",
      journal = {\nat},
         year = 2016,
        month = jan,
       volume = {529},
       number = {7586},
        pages = {364-367},
          doi = {10.1038/nature16171},
archivePrefix = {arXiv},
       eprint = {1601.00004},
 primaryClass = {astro-ph.SR},
       adsurl = {https://ui.adsabs.harvard.edu/abs/2016Natur.529..364S}
}

@ARTICLE{stello+16a,
       author = {{Stello}, Dennis and {Cantiello}, Matteo and {Fuller}, Jim and {Garcia}, Rafael A. and {Huber}, Daniel},
        title = "{Suppression of Quadrupole and Octupole Modes in Red Giants Observed by Kepler *}",
      journal = {\pasa},
         year = 2016,
        month = mar,
       volume = {33},
          eid = {e011},
        pages = {e011},
          doi = {10.1017/pasa.2016.9},
archivePrefix = {arXiv},
       eprint = {1602.05193},
 primaryClass = {astro-ph.SR},
       adsurl = {https://ui.adsabs.harvard.edu/abs/2016PASA...33...11S}
}

@ARTICLE{borucki+10,
       author = {{Borucki}, William J. and {Koch}, David and {Basri}, Gibor and {Batalha}, Natalie and {Brown}, Timothy and {Caldwell}, Douglas and {Caldwell}, John and {Christensen-Dalsgaard}, J{\o}rgen and {Cochran}, William D. and {DeVore}, Edna and {Dunham}, Edward W. and {Dupree}, Andrea K. and {Gautier}, Thomas N. and {Geary}, John C. and {Gilliland}, Ronald and {Gould}, Alan and {Howell}, Steve B. and {Jenkins}, Jon M. and {Kondo}, Yoji and {Latham}, David W. and {Marcy}, Geoffrey W. and {Meibom}, S{\o}ren and {Kjeldsen}, Hans and {Lissauer}, Jack J. and {Monet}, David G. and {Morrison}, David and {Sasselov}, Dimitar and {Tarter}, Jill and {Boss}, Alan and {Brownlee}, Don and {Owen}, Toby and {Buzasi}, Derek and {Charbonneau}, David and {Doyle}, Laurance and {Fortney}, Jonathan and {Ford}, Eric B. and {Holman}, Matthew J. and {Seager}, Sara and {Steffen}, Jason H. and {Welsh}, William F. and {Rowe}, Jason and {Anderson}, Howard and {Buchhave}, Lars and {Ciardi}, David and {Walkowicz}, Lucianne and {Sherry}, William and {Horch}, Elliott and {Isaacson}, Howard and {Everett}, Mark E. and {Fischer}, Debra and {Torres}, Guillermo and {Johnson}, John Asher and {Endl}, Michael and {MacQueen}, Phillip and {Bryson}, Stephen T. and {Dotson}, Jessie and {Haas}, Michael and {Kolodziejczak}, Jeffrey and {Van Cleve}, Jeffrey and {Chandrasekaran}, Hema and {Twicken}, Joseph D. and {Quintana}, Elisa V. and {Clarke}, Bruce D. and {Allen}, Christopher and {Li}, Jie and {Wu}, Haley and {Tenenbaum}, Peter and {Verner}, Ekaterina and {Bruhweiler}, Frederick and {Barnes}, Jason and {Prsa}, Andrej},
        title = "{Kepler Planet-Detection Mission: Introduction and First Results}",
      journal = {Science},
         year = 2010,
        month = feb,
       volume = {327},
       number = {5968},
        pages = {977},
          doi = {10.1126/science.1185402},
       adsurl = {https://ui.adsabs.harvard.edu/abs/2010Sci...327..977B}
}

@ARTICLE{mosser+12b,
       author = {{Mosser}, B. and {Goupil}, M.~J. and {Belkacem}, K. and {Marques}, J.~P. and {Beck}, P.~G. and {Bloemen}, S. and {De Ridder}, J. and {Barban}, C. and {Deheuvels}, S. and {Elsworth}, Y. and {Hekker}, S. and {Kallinger}, T. and {Ouazzani}, R.~M. and {Pinsonneault}, M. and {Samadi}, R. and {Stello}, D. and {Garc{\'\i}a}, R.~A. and {Klaus}, T.~C. and {Li}, J. and {Mathur}, S. and {Morris}, R.~L.},
        title = "{Spin down of the core rotation in red giants}",
      journal = {\aap},
         year = 2012,
        month = dec,
       volume = {548},
          eid = {A10},
        pages = {A10},
          doi = {10.1051/0004-6361/201220106},
archivePrefix = {arXiv},
       eprint = {1209.3336},
 primaryClass = {astro-ph.SR},
       adsurl = {https://ui.adsabs.harvard.edu/abs/2012A&A...548A..10M}
}

@ARTICLE{li+23,
       author = {{Li}, Gang and {Deheuvels}, S{\'e}bastien and {Li}, Tanda and {Ballot}, J{\'e}r{\^o}me and {Ligni{\`e}res}, Fran{\c{c}}ois},
        title = "{Internal magnetic fields in 13 red giants detected by asteroseismology}",
      journal = {\aap},
         year = 2023,
        month = dec,
       volume = {680},
          eid = {A26},
        pages = {A26},
          doi = {10.1051/0004-6361/202347260},
archivePrefix = {arXiv},
       eprint = {2309.13756},
 primaryClass = {astro-ph.SR},
       adsurl = {https://ui.adsabs.harvard.edu/abs/2023A&A...680A..26L}
}

@ARTICLE{deheuvels+23,
       author = {{Deheuvels}, S. and {Li}, G. and {Ballot}, J. and {Ligni{\`e}res}, F.},
        title = "{Strong magnetic fields detected in the cores of 11 red giant stars using gravity-mode period spacings}",
      journal = {\aap},
         year = 2023,
        month = feb,
       volume = {670},
          eid = {L16},
        pages = {L16},
          doi = {10.1051/0004-6361/202245282},
archivePrefix = {arXiv},
       eprint = {2301.01308},
 primaryClass = {astro-ph.SR},
       adsurl = {https://ui.adsabs.harvard.edu/abs/2023A&A...670L..16D}
}

@ARTICLE{li+22,
       author = {{Li}, Gang and {Deheuvels}, S{\'e}bastien and {Ballot}, J{\'e}r{\^o}me and {Ligni{\`e}res}, Fran{\c{c}}ois},
        title = "{Magnetic fields of 30 to 100 kG in the cores of red giant stars}",
      journal = {\nat},
         year = 2022,
        month = oct,
       volume = {610},
       number = {7930},
        pages = {43-46},
          doi = {10.1038/s41586-022-05176-0},
archivePrefix = {arXiv},
       eprint = {2208.09487},
 primaryClass = {astro-ph.SR},
       adsurl = {https://ui.adsabs.harvard.edu/abs/2022Natur.610...43L}
}

@ARTICLE{mosser+17,
       author = {{Mosser}, B. and {Belkacem}, K. and {Pin{\c{c}}on}, C. and {Takata}, M. and {Vrard}, M. and {Barban}, C. and {Goupil}, M. -J. and {Kallinger}, T. and {Samadi}, R.},
        title = "{Dipole modes with depressed amplitudes in red giants are mixed modes}",
      journal = {\aap},
         year = 2017,
        month = feb,
       volume = {598},
          eid = {A62},
        pages = {A62},
          doi = {10.1051/0004-6361/201629494},
archivePrefix = {arXiv},
       eprint = {1610.03872},
 primaryClass = {astro-ph.SR},
       adsurl = {https://ui.adsabs.harvard.edu/abs/2017A&A...598A..62M}
}

@ARTICLE{dupret+09,
       author = {{Dupret}, M. -A. and {Belkacem}, K. and {Samadi}, R. and {Montalban}, J. and {Moreira}, O. and {Miglio}, A. and {Godart}, M. and {Ventura}, P. and {Ludwig}, H. -G. and {Grigahc{\`e}ne}, A. and {Goupil}, M. -J. and {Noels}, A. and {Caffau}, E.},
        title = "{Theoretical amplitudes and lifetimes of non-radial solar-like oscillations in red giants}",
      journal = {\aap},
         year = 2009,
        month = oct,
       volume = {506},
       number = {1},
        pages = {57-67},
          doi = {10.1051/0004-6361/200911713},
archivePrefix = {arXiv},
       eprint = {0906.3951},
 primaryClass = {astro-ph.SR},
       adsurl = {https://ui.adsabs.harvard.edu/abs/2009A&A...506...57D}
}

@ARTICLE{fuller+15,
       author = {{Fuller}, Jim and {Cantiello}, Matteo and {Stello}, Dennis and {Garcia}, Rafael A. and {Bildsten}, Lars},
        title = "{Asteroseismology can reveal strong internal magnetic fields in red giant stars}",
      journal = {Science},
         year = 2015,
        month = oct,
       volume = {350},
       number = {6259},
        pages = {423-426},
          doi = {10.1126/science.aac6933},
archivePrefix = {arXiv},
       eprint = {1510.06960},
 primaryClass = {astro-ph.SR},
       adsurl = {https://ui.adsabs.harvard.edu/abs/2015Sci...350..423F}
}

@ARTICLE{ballot+11,
       author = {{Ballot}, J. and {Barban}, C. and {van't Veer-Menneret}, C.},
        title = "{Visibilities and bolometric corrections for stellar oscillation modes observed by Kepler}",
      journal = {\aap},
         year = 2011,
        month = jul,
       volume = {531},
          eid = {A124},
        pages = {A124},
          doi = {10.1051/0004-6361/201016230},
archivePrefix = {arXiv},
       eprint = {1105.4557},
 primaryClass = {astro-ph.SR},
       adsurl = {https://ui.adsabs.harvard.edu/abs/2011A&A...531A.124B}
}

@ARTICLE{mosser+12,
       author = {{Mosser}, B. and {Elsworth}, Y. and {Hekker}, S. and {Huber}, D. and {Kallinger}, T. and {Mathur}, S. and {Belkacem}, K. and {Goupil}, M.~J. and {Samadi}, R. and {Barban}, C. and {Bedding}, T.~R. and {Chaplin}, W.~J. and {Garc{\'\i}a}, R.~A. and {Stello}, D. and {De Ridder}, J. and {Middour}, C.~K. and {Morris}, R.~L. and {Quintana}, E.~V.},
        title = "{Characterization of the power excess of solar-like oscillations in red giants with Kepler}",
      journal = {\aap},
         year = 2012,
        month = jan,
       volume = {537},
          eid = {A30},
        pages = {A30},
          doi = {10.1051/0004-6361/20111735210.1086/141952},
archivePrefix = {arXiv},
       eprint = {1110.0980},
 primaryClass = {astro-ph.SR},
       adsurl = {https://ui.adsabs.harvard.edu/abs/2012A&A...537A..30M}
}

@ARTICLE{hekker+17,
       author = {{Hekker}, S. and {Christensen-Dalsgaard}, J.},
        title = "{Giant star seismology}",
      journal = {\aapr},
         year = 2017,
        month = jun,
       volume = {25},
       number = {1},
          eid = {1},
        pages = {1},
          doi = {10.1007/s00159-017-0101-x},
archivePrefix = {arXiv},
       eprint = {1609.07487},
 primaryClass = {astro-ph.SR},
       adsurl = {https://ui.adsabs.harvard.edu/abs/2017A&ARv..25....1H}
}

@ARTICLE{bugnet22,
       author = {{Bugnet}, L.},
        title = "{Magnetic signatures on mixed-mode frequencies. II. Period spacings as a probe of the internal magnetism of red giants}",
      journal = {\aap},
         year = 2022,
        month = nov,
       volume = {667},
          eid = {A68},
        pages = {A68},
          doi = {10.1051/0004-6361/202243167},
archivePrefix = {arXiv},
       eprint = {2208.14954},
 primaryClass = {astro-ph.SR},
       adsurl = {https://ui.adsabs.harvard.edu/abs/2022A&A...667A..68B}
}

@ARTICLE{cantiello+16,
       author = {{Cantiello}, Matteo and {Fuller}, Jim and {Bildsten}, Lars},
        title = "{Asteroseismic Signatures of Evolving Internal Stellar Magnetic Fields}",
      journal = {\apj},
         year = 2016,
        month = jun,
       volume = {824},
       number = {1},
          eid = {14},
        pages = {14},
          doi = {10.3847/0004-637X/824/1/14},
archivePrefix = {arXiv},
       eprint = {1602.03056},
 primaryClass = {astro-ph.SR},
       adsurl = {https://ui.adsabs.harvard.edu/abs/2016ApJ...824...14C}
}

@ARTICLE{takata16b,
       author = {{Takata}, Masao},
        title = "{Physical formulation of mixed modes of stellar oscillations}",
      journal = {\pasj},
         year = 2016,
        month = dec,
       volume = {68},
       number = {6},
          eid = {91},
        pages = {91},
          doi = {10.1093/pasj/psw093},
       adsurl = {https://ui.adsabs.harvard.edu/abs/2016PASJ...68...91T}
}

@ARTICLE{takata16a,
       author = {{Takata}, Masao},
        title = "{Asymptotic analysis of dipolar mixed modes of oscillations in red giant stars}",
      journal = {\pasj},
         year = 2016,
        month = dec,
       volume = {68},
       number = {6},
          eid = {109},
        pages = {109},
          doi = {10.1093/pasj/psw104},
       adsurl = {https://ui.adsabs.harvard.edu/abs/2016PASJ...68..109T}
}

@ARTICLE{loi20a,
       author = {{Loi}, Shyeh Tjing},
      journal = {\mnras},
         year = {2020},
        month = {4},
       volume = {493},
       number = {4},
        pages = {5726-5742},
          doi = {10.1093/mnras/staa581},
archivePrefix = {arXiv},
       eprint = {2002.11130},
 primaryClass = {astro-ph.SR},
       adsurl = {https://ui.adsabs.harvard.edu/abs/2020MNRAS.493.5726L}
}

@ARTICLE{loi20b,
       author = {{Loi}, Shyeh Tjing},
      journal = {\mnras},
         year = {2020},
        month = {8},
       volume = {496},
       number = {3},
        pages = {3829-3840},
          doi = {10.1093/mnras/staa1823},
archivePrefix = {arXiv},
       eprint = {2006.08635},
 primaryClass = {astro-ph.SR},
       adsurl = {https://ui.adsabs.harvard.edu/abs/2020MNRAS.496.3829L}
}

\begin{appendix}

\section{Coupling strength} \label{app: coupling}

\begin{table}[]
    \caption{Overview of the assumptions made for the computation of the coupling strength.}
    \centering
    \begin{tabular}{cccc}
    \hline\hline
    $\ell$ & coupling regime & Cowling & condition \\
    \hline
    $0$ & - & - & - \\
    $1$ & strong & no & Eq. \eqref{eq: coupling regimes} \\
    $1$ & intermediate & - & Eq. \eqref{eq: coupling regimes} \\
    $1$ & weak & no & Eq. \eqref{eq: coupling regimes} \\
    $2$ & weak & yes & - \\
    \hline\hline
    \end{tabular}
    \tablefoot{In the strong coupling regime, we used the prescription of \citet{takata16a} to compute $q_\ell$, while we adopted the expression of \citet{shibahashi79} in the weak coupling regime. In the intermediate regime, we used a linear interpolation (see Appendix \ref{app: implementation coupling}). If the Cowling approximation is not employed, we utilize the modified characteristic frequencies $\Hat{S}_\ell$ and $\Hat{N}$ instead of the regular $S_\ell$ and $N$ \citep{takata05,takata06a,takata16a}.}
    \label{tab: overview coupling}
\end{table}

The coupling strength, $q_\ell$, of the mixed modes of the spherical degree $\ell$ is a measure for the amount of interaction between the p- and the g-mode cavities. It can theoretically take values between 0 (no interaction) and 1 (maximum interaction). The two cavities are separated by the evanescent zone, in which the oscillations are damped exponentially. In red giant stars, part of the energy of the oscillations can be exchanged between the two cavities by transmission through the evanescent zone \citep[e.g.,][]{shibahashi79, unno+89, aerts+10book, hekker+17}. Therefore, $q_\ell$ is closely connect to the modulus of the transmission coefficient, $|T|$, associated with the evanescent zone, which generally depends on the thickness and the stratification of the evanescent zone \citep[e.g.,][]{shibahashi79, unno+89, takata16a}.

In this section, we describe the calculation of the exponent $X$, which can be used to estimate $|T|$ (and thus $q_\ell$) for a given stellar model via
\begin{gather}
    |T| = e^{-\pi X}.
    \label{eq: transmission X}
\end{gather}
Note that both $|T|$ and $X$ depend on $\nu$ and $\ell$. Using asymptotic theory, expressions for $X$ were found in the limit of a wide evanescent zone \citep{shibahashi79} and in the limit of a very thin evanescent zone \citep{takata16a}. They are discussed in Appendix \ref{app: weak coupling} and \ref{app: strong coupling}, respectively. Before that, we introduce a handful of general parameters needed for the calculation of $X$ in Appendix \ref{app: genereal parameters coupling} and \ref{app: radial coordinate}.
Finally, we describe the regimes in which we apply the two limits to our evolutionary tracks in Appendix \ref{app: coupling regimes} and our procedure for computing $q_\ell$ in Appendix \ref{app: implementation coupling}.

\subsection{Characteristic frequencies} \label{app: genereal parameters coupling}

The Cowling approximation is a frequently used assumption for stellar oscillations and states that the influence of the disturbance of the gravitational potential can be neglected \citep{cowling41}. It reduces the fourth-order system of differential equations describing the propagation of multipolar adiabatic oscillations to a second-order system \citep[e.g.,][]{shibahashi79, unno+89}.
For the dipole modes, \citet{pincon+20} show that adopting the Cowling approximation can lead to an incorrect estimate of $q_\ell$. Fortunately, \citet{takata05, takata06a} demonstrated that the fourth-order governing system of the dipole modes can be reduced to second order without adopting the Cowling approximation. This is possible due to the geometric characteristics exclusive to the dipolar oscillations, which do not inherently conserve the momentum of each mass shell of the star \citep{takata05}. Therefore, the conservation of momentum of the entire star constitutes an additional constraint to the propagation of dipole modes, thus simplifying their treatment.

The effect of the perturbation of the gravitational potential can be taken into account by introducing a modification to the Lamb frequency $S_\ell$ and the buoyancy frequency $N$. The regular (unmodified) characteristic frequencies are given by
\begin{gather}
    S_\ell = \frac{\sqrt{\ell(\ell+1)}\ c_{\rm s}}{r}, \\
    N = \sqrt{ \frac{g}{r} \left( \frac{1}{\gamma} \frac{\text{d} \ln p}{\text{d} \ln r} - \frac{\text{d} \ln \rho}{\text{d} \ln r} \right) },
    \label{eq: buoyancy frequncy}
\end{gather}
where $c_{\rm s}$ is the adiabatic sound speed and $\gamma$ is the first adiabatic index.
The modified characteristic frequencies are defined as \citep{takata05, takata06a, takata16a}
\begin{gather}
    \Hat{S}_\ell = J S_\ell \quad \text{and} \quad \Hat{N} = \frac{N}{J},
    \label{eq: modified characteristic frequencies}
\end{gather}
with
\begin{gather}
    J = 1 - \frac{4 \pi r^3 \rho}{3 m},
\end{gather}
where $m$ is the mass coordinate. The Cowling approximation corresponds to the case $J = 1$.

Since a relaxation of the Cowling approximation as utilized in Eq. \eqref{eq: modified characteristic frequencies} is only possible for the dipole modes, we use the modified characteristic frequencies $\Hat{S}_\ell$ and $\Hat{N}$ for the calculation of $q_{\ell=1}$ and the regular $S_\ell$ and $N$ for the calculation of $q_{\ell=2}$. In Table \ref{tab: overview coupling} we summarize our assumptions for computing the coupling strength for different $\ell$.
In the following, we focus our discussion on the modified frequencies $\Hat{S}_\ell$ and $\Hat{N}$. However, the procedure for calculating $q_\ell$ in the weak coupling regime (see Appendix \ref{app: weak coupling}) can also be applied to the quadrupole modes. To obtain the corresponding expressions, it is sufficient to simply replace the modified frequencies with the regular ones.

\subsection{Parameters $\mathfrak{P}$ and $\mathfrak{Q}$} \label{app: radial coordinate}

It is convenient to make a change in variables from $r$ to $s$, such that \citep{takata16a}
\begin{gather}
    s = \ln r - \frac{1}{2} (\ln r_1 + \ln r_2).
\end{gather}
Here, $r_1$ is defined as the radius at which $\Hat{S}_\ell$ is equal to the angular frequency of the oscillations $\omega = 2\pi\nu$. Likewise, $r_2$ is the radius at which $\Hat{N} = \omega$.
The center of the evanescent zone is located at $s = 0$ and its boundaries are located at $-|s_0|$ and $|s_0|$, where
\begin{gather}
    s_0 = \frac{1}{2} (\ln r_1 - \ln r_2).
\end{gather}
In addition, we follow \citet{takata16a} and introduce the functions $\mathfrak{P}$ and $\mathfrak{Q}$ as
\begin{gather}
    \mathfrak{P} = \ell(\ell+1)\ J \left( 1- \frac{\omega^2}{\Hat{S}_\ell^2} \right), \\
    \mathfrak{Q} = J \left( 1 - \frac{\Hat{N}^2}{\omega^2} \right),
\end{gather}
where $r_1$ is the zero crossing of $\mathfrak{P}$ and $r_2$ is the zero crossing of $\mathfrak{Q}$.

\subsection{Weak coupling} \label{app: weak coupling}

The weak coupling regime corresponds to the limit of a wide evanescent zone. In this regime, the exponent $X$ in Eq. \eqref{eq: transmission X} is made up of a single term that depends on the thickness of the evanescent zone:
\begin{gather}
    X = X_{\rm width}.
\end{gather}
It is given by the integral of the wavenumber over the evanescent zone \citep{shibahashi79, takata16a}, such that
\begin{gather}
    X_{\rm width} = \frac{1}{\pi} \int_{-|s_0|}^{|s_0|} \sqrt{ \mathfrak{P}\mathfrak{Q}}\ {\rm d}s.
    \label{eq: X thickness}
\end{gather}
The wider the evanescent zone, the larger $X_{\rm width}$, causing $q_\ell$ to decrease. In the regime of a wide evanescent zone, the relationship between the coupling strength and the transmission coefficient is \citep{shibahashi79}
\begin{gather}
    q_\ell = \frac{|T|^2}{4}.
    \label{eq: transmission -> coupling weak}
\end{gather}

\subsection{Strong coupling} \label{app: strong coupling}

The strong coupling regime corresponds to the limit of a very thin evanescent zone. Here, an additional contribution to $X$ accounting for the stratification of the evanescent zone arises \citep{takata16a}:
\begin{gather}
    X = X_{\rm width} + X_{\rm grad}.
\end{gather}
The stratification term depends on the gradient of the equilibrium quantities in the evanescent zone and is evaluated at its center. It can be expressed as \citep{takata16a, vanRossem+24}
\begin{gather}
    X_{\rm grad} = \left( \frac{s_0}{2 \sqrt{\mathfrak{P}\mathfrak{Q}}} \left[ \frac{1}{4} \frac{\text{d}}{\text{d}s} \left(  \ln\frac{\mathfrak{P}(s + s_0)}{\mathfrak{Q}(s_0 - s)} \right) \right. \right. \notag\\
    \qquad\qquad\qquad \left.\left. - \frac{J}{2} \left(\frac{2\ g}{\Hat{S}_\ell^2\ e^{s_0}} - \frac{\Hat{N}^2\ e^{s_0}}{g} - 1\right) \right] ^2 \right)_{s=0}.
    \label{eq: X stratification}
\end{gather}

The relationship between the coupling strength and the transmission coefficient is given by \citep{takata16a}
\begin{gather}
    q_\ell = \frac{1 - \sqrt{1 - |T|^2}}{1 + \sqrt{1 - |T|^2}}.
    \label{eq: transmission -> coupling}
\end{gather}
It is apparent that Eq. \eqref{eq: transmission -> coupling} approaches Eq. \eqref{eq: transmission -> coupling weak} for small $|T|$. However, it has been demonstrated by \citet{pincon+20} that $X_{\rm grad}$ does not necessarily vanish when increasing the width of the evanescent zone. For a broadening evanescent zone, there is thus no smooth transition between the values of $q_\ell$ obtained in the strong and weak coupling regime. Such a discontinuity is most likely unphysical \citep[e.g.,][]{mosser+17coupling,vanLier+25} and would significantly impact the mixed mode visibilities. To mitigate this, we introduce an empirical transition between the two regimes in the next section.

\subsection{Coupling regimes on the red-giant branch} \label{app: coupling regimes}

\subsubsection{Dipole modes}

During the evolution of an RGB star, the evanescent zone undergoes three phases. At first, it is completely located inside the radiative core of the star. Afterwards, during the second phase, the evanescent zone partly moves into the convective outer layer. In this phase, it is thus characterized by being partly radiative and convective. In the third phase, the evanescent zone is located entirely in the lower part of the outer convective envelope of the star. 

The importance of these phases for the estimation of $q_{\ell=1}$ has been highlighted by several studies \citep{pincon+19, pincon+20, jiang+22, vanRossem+24, vanLier+25}.
For the dipole modes, these studies suggest that $q_{\ell=1}$ can be described by the strong coupling prescription in the first phase and by the weak coupling prescription in the third phase. 
Thus, the transition from the strong to the weak coupling regime is expected to take place during the second phase, where the evanecent zone is partly radiative and convective. However, recent empirical work by \citet{vanLier+25} suggests that the weak coupling prescription actually approximates the behavior of $q_{\ell=1}$ with reasonable accuracy in the second phase \citep[see also][]{pincon+20}. This is caused by a fortunate cancellation of the errors introduced by two separate approximations, which are most likely not valid in that regime, namely the usage of Eq. \eqref{eq: transmission -> coupling weak} instead of Eq. \eqref{eq: transmission -> coupling} and the assumption that $X_{\rm grad} = 0$ \citep[for details, see][]{vanLier+25}.

An additional complication arising in the second phase is that the profile of the buoyancy frequency features a narrow spike located close to the base of the outer convection zone. This spike corresponds to a discontinuity in the chemical composition of the star and is a remnant of the first dredge-up. If the spike is located between $r_1$ and $r_2$, it separates the evanescent zone into two parts, provided that the value of the buoyancy frequency at the tip of the spike is equal or larger than the selected oscillation frequency (i.e., if $\omega \leq \max(\Hat{N}(s))_{|s|\leq|s_0|}$). This breaks the assumption that there is only one evanescent zone. 
Moreover, the spike constitutes a glitch, because the stellar structure locally changes on a scale smaller than the wavelength of the oscillations \citep[e.g.,][]{gough90, cunha+15, jiang+22, vanLier+25}, thus violating the asymptotic approximation, which is essential for both coupling prescriptions. 
Furthermore, the presence of the spike in the evanescent zone affects the gradients required for determining $X_{\rm grad}$, which complicates the calculation.

In this work, we do not calculate $q_{\ell=1}$ while the spike in the buoyancy frequency separates the evanescent zone into two parts. Instead, we use a linear interpolation between the strong and weak coupling regimes as a crude estimate of the coupling strength (see Appendix \ref{app: implementation coupling}), so that $q_{\ell=1}$ behaves coherently throughout the evolution on the RGB and is not subject to sudden discontinuous variations.
The coupling regimes adopted for the determination of $q_{\ell=1}$ can be summarized as follows:
\begin{gather}
    q_{\ell=1}:
    \begin{cases}
      \text{(i)\ \ \ interpolation}, & \text{if}\ \omega \leq \max[\Hat{N}(s)]_{|s|\leq|s_0|},\\
      \text{(ii)\ \ strong}, & \text{if not (i) and}\ |s_0| < s_{\rm conv},\\
      \text{(iii)\ weak}, & \text{if not (i) and}\ |s_0| \geq s_{\rm conv}.
      \label{eq: coupling regimes}
    \end{cases}
\end{gather}
Here, $s_{\rm conv}$ denotes the location of the lower boundary of the convective envelope of the star. The resulting values of $q_{\ell=1}$ transition smoothly from the strong to the weak coupling regime and are consistent with estimates from previous studies \citep[e.g.,][]{vanLier+25}.

\subsubsection{Quadrupole modes}

The Lamb frequency of the quadrupole modes $S_{\ell=2}$ is systematically larger than $S_{\ell=1}$. For a given frequency $\nu$, the evanescent zone is thus wider for the quadrupole modes compared to the dipole modes. Therefore, we assume that the coupling of the quadrupole modes always falls into the weak coupling regime, regardless of the width the evanescent zone, and even if the spike of the buoyancy frequency separates the evanescent zone. Since \citet{vanLier+25} show that the presence of the spike does not strongly affect the estimation of $X_{\rm width}$, we do not expect a strong influence on our $q_{\ell=2}$ estimate.

\subsection{Implementation} \label{app: implementation coupling}

\begin{figure}[]
    \centering
    \resizebox{\hsize}{!}{\includegraphics{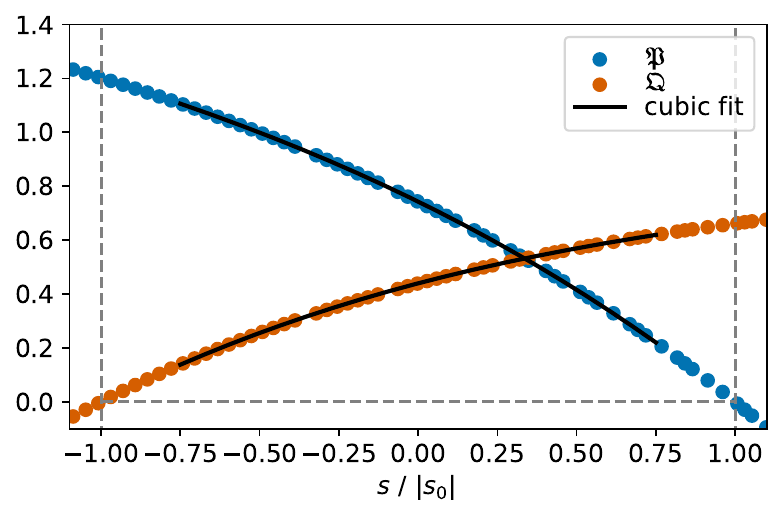}}
    \caption{Parameters $\mathfrak{P}$ and $\mathfrak{Q}$ as a function of $s$ for a model with $M_\star = 1.25\ M_\odot$ and $\nu_{\rm max} \approx 250$ $\mu$Hz. The oscillation frequency is $\omega = 2\pi\nu_{\rm max}$ and the spherical degree is $\ell=1$. The evanescent zone is fully radiative. Colored dots indicate the value of $\mathfrak{P}$ (in blue) and $\mathfrak{Q}$ (in red) in each cell of the MESA grid. Solid black lines indicate the corresponding cubic fits. Dashed gray lines indicate the boundaries of the evanescent zone (vertical) and the zero line (horizontal). This figure is representative for the behavior of $\mathfrak{P}$, $\mathfrak{Q}$, and the fitted polynomials for all models and frequencies if condition (i) in Eq. \eqref{eq: coupling regimes} is fulfilled.}
    \label{fig: coupling P and Q}
\end{figure}

\begin{figure}[]
    \centering
    \resizebox{\hsize}{!}{\includegraphics{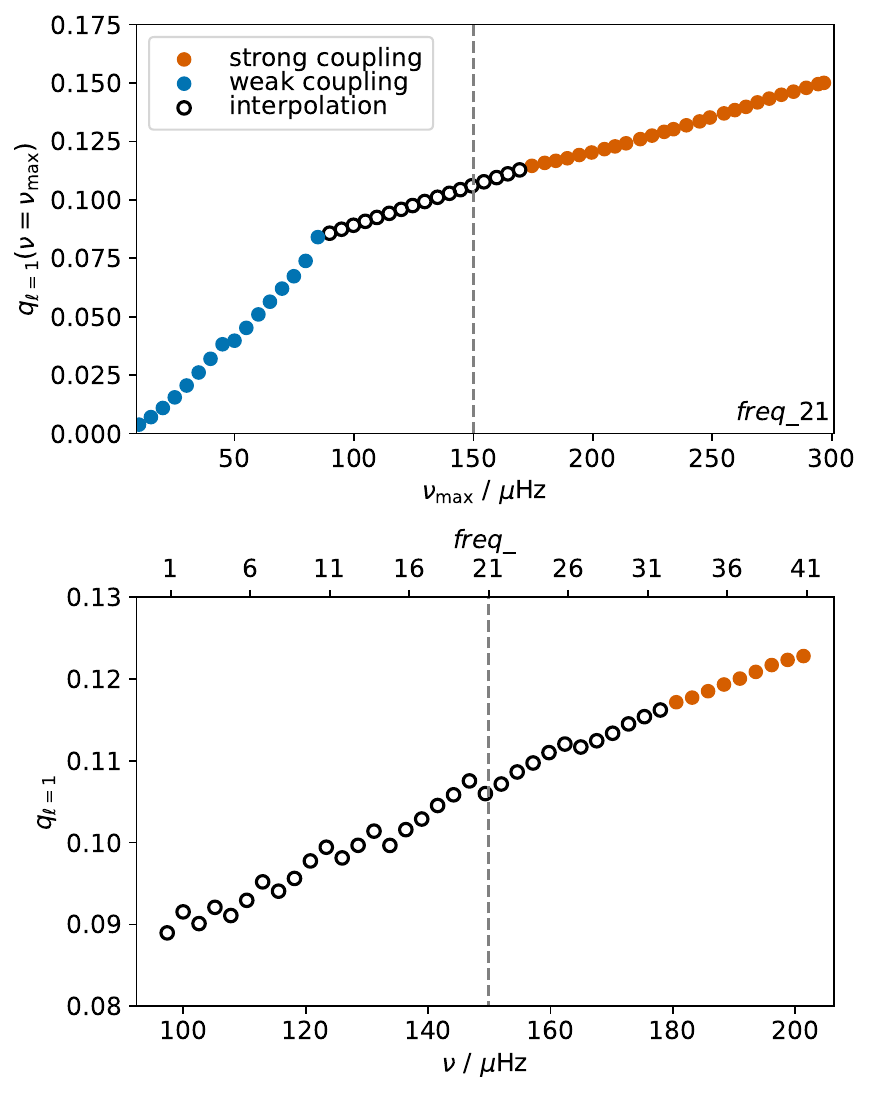}}
    \caption{\textit{Top:} Evolution of the coupling strength of the dipole modes at the frequency point \textit{freq\_21} (see Appendix \ref{app: implementation coupling}) as a function of $\nu_{\rm max}$ for the evolutionary track with $M_\star = 1.25\ M_\odot$. The frequency point \textit{freq\_21} is by definition always located at $\nu_{\rm max}$ of each model. The coupling strengths calculated using the weak (strong) coupling prescription are shown in blue (red). Black circles indicate the coupling strengths estimated using the interpolation procedure described in Appendix \ref{app: implementation coupling}. The gray dashed line marks the location of the model with $\nu_{\rm max} \approx 150$ $\mu$Hz. \textit{Bottom:} Coupling strength of the dipole modes as a function of frequency for the model with $\nu_{\rm max} \approx 150$ $\mu$Hz. We also indicate the frequency points \textit{freq\_1} to \textit{freq\_41} in the top axis. The colors are the same as above. The gray dashed line marks the location of the frequency point \textit{freq\_21}, causing the values of $q_{\ell=1}$ intersecting with the gray dashed lines to be equal in the top and bottom panel.}
    \label{fig: coupling freq_21}
\end{figure}

\begin{figure}[]
    \centering
    \resizebox{\hsize}{!}{\includegraphics{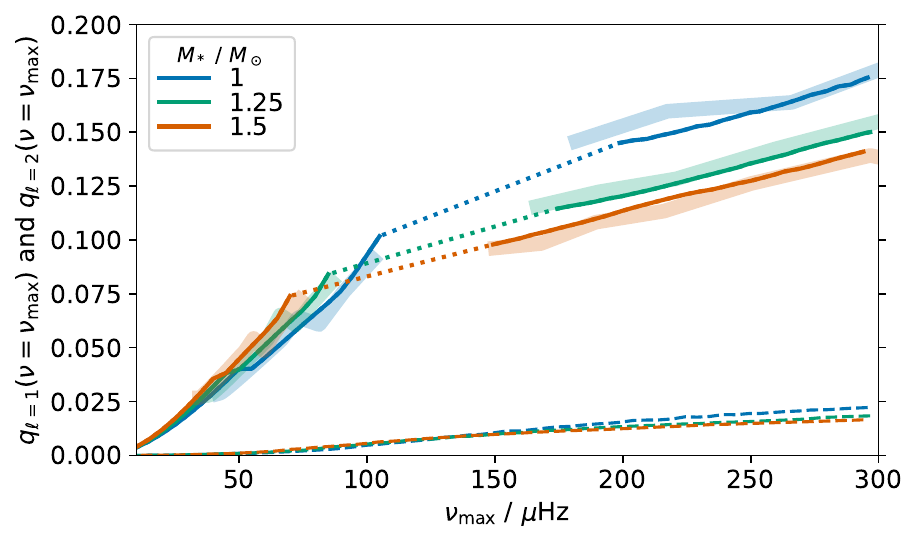}}
    \caption{Coupling strength of the dipole and quadrupole modes at $\nu=\nu_{\rm max}$ as a function of the frequency of maximum oscillation power. Colors indicate different stellar masses. For each mass, the coupling strength of the quadrupole modes is represented by the dashed line. The coupling strength of the dipole modes is shown as a solid line when it was determined using the weak or strong coupling prescriptions at low or high values of $\nu_{\rm max}$, respectively, and as a dotted line when it was estimated using an interpolation (see Eq. \eqref{eq: coupling regimes}). The thick transparent lines show the coupling strength calculated by \citet{vanLier+25} for their models of the respective mass using the same coupling prescription as for the corresponding solid line. We have checked that the differences are mainly due to the fact that \citet{vanLier+25} used a low artificial diffusion instead of an exponential overshoot to smooth the stellar structure.}
    \label{fig: evolution coupling}
\end{figure}

The coupling strength is computed for a given stellar model, frequency, and spherical degree. For each model, we calculate $q_{\ell}$ for 41 frequencies that are evenly distributed on the interval $\nu_{\rm max} - 4 \Delta\nu \leq \frac{\omega}{2\pi} \leq \nu_{\rm max} + 4 \Delta\nu$.
As a first step, we compute the value of $\mathfrak{P}$ and $\mathfrak{Q}$ in each grid cell of the MESA profiles and find the zero-crossings corresponding to $r_1$ and $r_2$ using a linear interpolation between the grid cells. Then, we determine $s_0$. For the dipole modes, we also determine $s_{\rm conv}$ and check which coupling regime is appropriate using the conditions in Eq. \eqref{eq: coupling regimes}. Depending on the regime, we calculate either both $X_{\rm width}$ and $X_{\rm grad}$, only $X_{\rm width}$, or neither.

If only $X_{\rm width}$ is computed (i.e., if condition (iii) is fulfilled), we use a trapezoidal integration to solve the integral in Eq. \eqref{eq: X thickness}.
If both $X_{\rm width}$ and $X_{\rm grad}$ are computed (i.e., if condition (ii) is fulfilled), we first fit the values of $\mathfrak{P}$ and $\mathfrak{Q}$ around the center of the evanescent zone (i.e., for $-0.75 |s_0| \leq s \leq 0.75 |s_0|$) using a cubic polynomial. This approximation is justified if the evanescent zone lies within the radiative core and the spike in the buoyancy frequency is not located within the evanescent zone (see Fig. \ref{fig: coupling P and Q}). 
The polynomials are used instead of $\mathfrak{P}$ and $\mathfrak{Q}$ to improve the stability of the subsequent calculation of the gradients, which can otherwise be prone to numerical fluctuations due to the calculation of logarithmic derivatives on a finite grid.
Then, we determine $X_{\rm width}$ using Eq. \eqref{eq: X thickness} and $X_{\rm grad}$ using Eq. \eqref{eq: X stratification}.

If neither $X_{\rm width}$ nor $X_{\rm grad}$ are calculated (i.e., if condition (i) is fulfilled), we first estimate $q_{\ell=1}$ using Eqs. \eqref{eq: X thickness} and \eqref{eq: X stratification} for each of our predefined 41 frequencies of all stellar models of an evolutionary track where this is possible. Then, we label the points of the aforementioned frequency grid for each stellar model with increasing frequency from \textit{freq\_1} to \textit{freq\_41}. For each frequency point \textit{freq\_}, we thus obtain the evolution of $q_{\ell=1}$ at that frequency point as a function of the evolution on the RGB\footnote{Note that the frequency $\omega$ corresponding to an individual frequency point \textit{freq\_} changes while the star evolves and is therefore different for each stellar model (see upper panel of Fig. \ref{fig: coupling freq_21}).}.
This is shown in the upper panel of Fig. \ref{fig: coupling freq_21} for the frequency point \textit{freq\_21} as an example. We obtain a similar behavior for the other frequency points \textit{freq\_}. In the upper panel of Fig. \ref{fig: coupling freq_21}, we can see three phases.
At an early or late stage in the evolution on the RGB, conditions (ii) or (iii) in Eq. \eqref{eq: coupling regimes} is satisfied and $q_{\ell=1}$ can be calculated using the asymptotic expressions. Between those two regimes, there is a region where $q_{\ell=1}$ has not been determined, because condition (i) is fulfilled. In this intermediate region, we adopt a linear interpolation as a function of frequency to estimate $q_{\ell=1}$, such that we obtain a value for the coupling strength at each frequency point \textit{freq\_} for each stellar model.
After the interpolation in time has been carried out for all frequency points \textit{freq\_}, the resulting values of $q_{\ell=1}$ are cast back onto the models from which they originate. We thus obtain 41 estimates of $q_{\ell=1}$ in the interval $\nu_{\rm max} - 4 \Delta\nu \leq \nu \leq \nu_{\rm max} + 4 \Delta\nu$ for each model. An example of this is shown in the lower panel of Fig. \ref{fig: coupling freq_21}. The range of calculated values of $q_{\ell}$ covers the range of observable frequencies for each model and can be interpolated to provide an estimate of the coupling strength of the oscillations at each frequency $\nu$ within that interval.
The resulting values of $q_\ell$ are shown in Fig. \ref{fig: evolution coupling} for our three evolutionary tracks.

\end{appendix}
\end{document}